\PassOptionsToPackage{breaklinks, colorlinks = true, linkcolor = blue, urlcolor = blue, citecolor = magenta, anchorcolor = blue}{hyperref}
\documentclass[
pre,twocolumn,notitlepage,longbibliography,amsmath,amssymb,floats,superscriptaddress,nofootinbib,10pt]{revtex4-2}
\usepackage{graphicx,color}
\usepackage{amsmath}
\usepackage{amsfonts, amsbsy}
\usepackage{amssymb}
\usepackage{eqnarray}
\usepackage{bm}
\usepackage{blkarray}
\usepackage{mathtools}
\usepackage{array}
\usepackage{psfrag}
\usepackage{multirow}
\usepackage{textcomp}
\usepackage{units}
\usepackage{lipsum}
\usepackage[title]{appendix}
\usepackage{soul}
\usepackage{titlesec}
\usepackage{times}
\usepackage{enumitem}
\usepackage{comment}
\usepackage{braket}
\usepackage[sort&compress]{natbib}
\usepackage{orcidlink}
 
\newcommand{\mcE}{\mathcal{E}}	
	
	\newcommand{\mcJ}{\mathcal{J}}
	\newcommand{\mcL}{\mathcal{L}}
\newcommand{\mcM}{\mathcal{M}}	
	\newcommand{\mcP}{\mathcal{P}}
	
	\newcommand{\mcT}{\mathcal{T}}

	\newcommand{\bmD}{\boldsymbol{D}}
	
\newcommand{\bmG}{\boldsymbol{G}}

	\newcommand{\bmT}{\boldsymbol{T}}

	\newcommand{\mfP}{\mathfrak{P}}
	
\newcommand{\mfS}{\mathfrak{S}}

	\newcommand{\mfv}{\mathfrak{v}}

\newcommand{\ovg}{\overline{g}}	\newcommand{\ovh}{\overline{h}}

	\newcommand{\ovv}{\overline{v}}

\newcommand{\tdA}{\tilde{A}}

	\newcommand{\tdh}{\tilde{h}}

\newcommand{\Lm}{\Lambda}
\newcommand{\dl}{\delta}

\newcommand{\gm}{\gamma}
\newcommand{\Gm}{\Gamma}

\newcommand{\lt}{\left}
\newcommand{\rt}{\right}
\newcommand{\mass}{M} 

\usepackage[breaklinks, colorlinks = true, linkcolor = blue, urlcolor = blue, citecolor = magenta, anchorcolor = blue]{hyperref}
\usepackage{float}

\usepackage{xcolor}

\begin{document}
\title{Curvature Gradients Turn Shear into Superflow in Supersolids}
 
\author{Ygor de Oliveira Souza\orcidlink{0009-0003-1001-2822}}
\email{ygor.deoliveira-souza@pwr.edu.pl}
\affiliation{Institute of Theoretical Physics, Wroc\l{}aw University of Science and Technology, 50-370 Wroc\l{}aw, Poland}
\author{Piotr Sur\'owka\orcidlink{0000-0003-2204-9422}}
\email{piotr.surowka@pwr.edu.pl}
\affiliation{Institute of Theoretical Physics, Wroc\l{}aw University of Science and Technology, 50-370 Wroc\l{}aw, Poland}

\date{\today}

\begin{abstract}
Supersolids on curved surfaces are becoming experimentally accessible. We develop a Galilean-covariant effective field theory of supersolids on curved surfaces and show that gradients of Gaussian curvature couple lattice shear to compression with a strength fixed by the shear modulus and the curvature gradient. As a result, shear drives superflow. On spheroids this yields geometric selection rules and multiplet splittings; near mode crossings, weak deformations produce hybrids in which superfluid and lattice carry comparable kinetic energy, and shear becomes visible as a density response.
\end{abstract}
 
\maketitle

{\it Introduction}$-$ Supersolids combine two forms of order conventionally associated with
distinct phases of matter: the rigidity of a crystal and the phase
coherence of a superfluid
\cite{Andreev1969,Chester1970,Leggett1970,Recati2023Supersolidity}.
Microscopically, this coexistence is reflected in the simultaneous
breaking of translational and global \(U(1)\) symmetries, giving rise to
coupled lattice and superfluid degrees of freedom
\cite{Recati2023Supersolidity}. Originally proposed in the context of
solid helium \cite{Andreev1969,Chester1970,Leggett1970}, supersolidity
has now been realized in several ultracold-atom platforms, including
spin--orbit-coupled condensates \cite{Li2017Stripe}, quantum gases
coupled to optical cavities \cite{Leonard2017Supersolid}, and, most
prominently, dipolar quantum gases, where coherent density-modulated
states can be prepared and controlled
\cite{Bottcher2019Transient,Tanzi2019Observation,
Chomaz2019LongLived,Norcia2021TwoDimensional}.

Recent developments have considerably broadened the phenomenology of
supersolids. In dipolar quantum gases, experiments have progressed from
observing coherent density-modulated states
\cite{Bottcher2019Transient,Tanzi2019Observation,
Chomaz2019LongLived,Norcia2021TwoDimensional} and their characteristic
Goldstone modes \cite{Tanzi2019SymmetryBreaking,Guo2019Goldstone} to
probing genuinely hydrodynamic signatures, including quantized vortices
and the response of supersolids to rotation \cite{Casotti2024Vortices}.
These systems realize cluster supersolidity through the roton-driven
modulation of an initially superfluid condensate and offer exceptional
control over interactions, dimensionality, and confinement
\cite{Chomaz2019LongLived,Norcia2021TwoDimensional,
Recati2023Supersolidity}.

A complementary mechanism was recently identified in ion-doped helium
nanodroplets, where the polarization field of a central ion produces a
crystallized ``snowball'' surrounded by superfluid helium
\cite{AcostaMatos2025Transitional}. Between these regions, a supersolid
layer with icosahedral density modulation can emerge on an approximately
spherical surface. This result demonstrates that supersolid order can
arise in a finite curved geometry without a conventional roton
instability \cite{AcostaMatos2025Transitional}. More broadly, curved
geometries are becoming experimentally accessible in ultracold atomic
systems through ring-, shell-, and bubble-shaped traps
\cite{Tononi2023CurvedGases,Tononi2025AxiallySymmetric}. In particular, microgravity experiments
aboard the International Space Station have enabled the realization of
ultracold atomic bubbles by suppressing the gravitational sag that
prevents the formation of homogeneous closed shells on Earth
\cite{Carollo2022AtomicBubbles}. These developments motivate a general
description of supersolidity beyond flat, translationally invariant
settings, in which curvature and global geometry become intrinsic
ingredients of the many-body problem.

Here we develop a Galilean-covariant effective field theory of supersolids on curved surfaces, describing superfluid order by its \(U(1)\) phase and crystalline order by material-coordinate fields. In flat space, longitudinal lattice excitations hybridize with the superfluid mode, while transverse shear remains decoupled. We show that uniform curvature preserves this separation, whereas curvature gradients couple shear to compression and thereby to superfluid flow. On spheroidal surfaces, this mechanism produces characteristic spectral splittings and avoided crossings, where even weak geometric deformations can generate strongly mixed shear and superfluid modes. Crucially, a shear excitation that carries no density response on a sphere becomes visible through resonant density oscillations on a spheroid, providing a direct experimental signature of curvature-induced superflow.

\textit{Effective theory.}---We describe the superfluid by its $U(1)$ phase $\theta$
and the crystal by material coordinates $X^A(t,\mathbf{x})$, $A=1,2$, which assign
to each point of the surface the label of the lattice site located there. Unlike
broken translations, which have no meaning on a generic surface, these labels are
defined on any geometry \cite{Nicolis2015Zoology}.
The surface enters only through its metric $g_{ab}$ and the associated covariant
derivative $\nabla_a$. Coupling the theory to a Newton--Cartan background makes
Galilean invariance manifest~\cite{Son2015,Jensen2018}: on a static surface it
requires $\theta$ and $X^A$ to appear only through
\begin{equation}
\begin{aligned}
\mu &= \dot{\theta}
      - \frac{(\nabla\theta)^2}{2\mass},\\
W^A &= -\dot{X}^A
      + \frac{1}{\mass}\nabla^a\theta\,\nabla_a X^A,\\
U^{AB} &= \nabla^a X^A\,\nabla_a X^B ,
\end{aligned}
\label{eq:invariants}
\end{equation}
with $\mass$ the particle mass. Namely these are the local chemical potential, the velocity of the lattice relative to the
superfluid, and the metric that the surface induces on the crystal. In flat space
these reduce to the variables of Son~\cite{Son2005SupersolidEFT}.

The low-energy action is $S=\int dt\,d^2x\sqrt{g}\,F$, with the pressure $F$ a
function of the invariants~\eqref{eq:invariants}. Elastic deformations are measured
by the strain $k_{AB}=\tfrac12(G_{AB}-U_{AB})$, where $G_{AB}$ is the metric of the
unstressed crystal. The key physical input is the choice of this reference state.
A supersolid that self-assembles on a curved surface relaxes to its own unstrained
configuration, so we assume that $G_{AB}$ is the equilibrium induced metric and inherits the
curvature of the substrate. The crystal is therefore a non-Euclidean
plate~\cite{non_euc_plates}, not a flat lattice forced onto a curved surface. Time
reversal and isotropy then fix $F$ to quadratic order,
\begin{equation}
\begin{aligned}
F \simeq{}&
P(\mu)+\frac{\rho_n}{2}W_AW^A
+2\lambda\,\delta\mu\,k^A{}_A\\
&-\frac{B}{2}\bigl(k^A{}_A\bigr)^2\\
&-\mu_e\left(
k_{AB}k^{AB}-\frac{1}{2}\bigl(k^A{}_A\bigr)^2
\right),
\end{aligned}
\label{eq:action}
\end{equation}
with indices raised by $G^{AB}$. Here $P'=n$ and $P''=\chi$ are the local density and
compressibility of the condensate, $\rho_n$ is the normal density, $\lambda$ couples
density to strain, and $B$ and $\mu_e$ are the bulk and shear moduli, all coeficcients assumed constant.
The superfluid stiffness is not an independent parameter: it receives contributions from both the
Galilean-invariant $\mu$ and the counterflow term, which combine into
$\rho_s=\mass n-\rho_n$. The mass current takes the two-fluid form
\begin{equation}
\mathbf{J}=\rho_s\,\mathbf{u}_s+\rho_n\,\mathbf{u}_X,
\label{eq:twofluid}
\end{equation}
with $\mathbf{u}_s=-\nabla\theta/\mass$ the superfluid velocity and $\mathbf{u}_X$ the
velocity of the lattice. In flat space the theory reproduces the supersolid
hydrodynamics of Refs.~\cite{Son2005SupersolidEFT,yoo_hydrodynamic_2010}; details, including the full set of
Ward identities, are given in the Supplemental Material (SM).

\begin{figure}[t]
\centering
\includegraphics[width=\columnwidth]{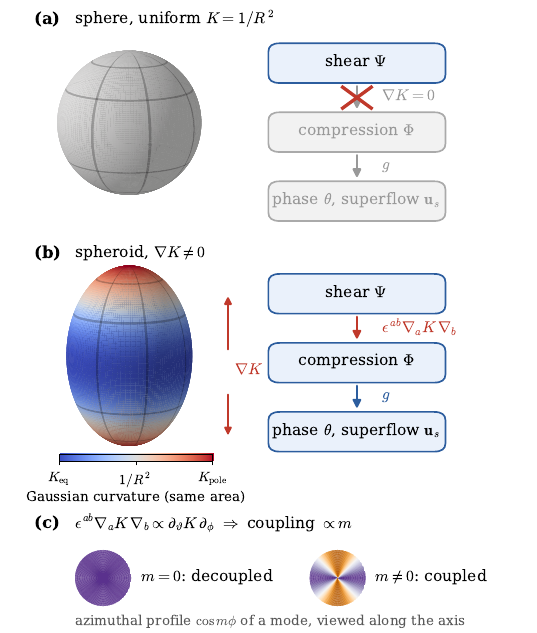}
\caption{Shear-driven superflow from curvature gradients.
(a)~On a sphere the Gaussian curvature $K$ is uniform; lattice shear $\Psi$
decouples from compression $\Phi$ and carries neither superflow nor density.
(b)~On a spheroid, $\nabla K\neq0$ couples shear to compression, which couples to
the superfluid phase $\theta$ through $g$, so a shear deformation drives a
superflow $\mathbf{u}_s$. Color: $K$. Because
$\epsilon^{ab}\nabla_aK\nabla_b\propto\partial_\vartheta K\,\partial_\phi$, the
coupling vanishes for azimuthally symmetric ($m=0$) modes.}
\label{fig:schematic}
\end{figure}

\textit{Dynamics on an arbitrary surface.}---We expand around equilibrium,
$\theta=\mu_0t+\delta\theta$ and $X^A=X^A_0+\delta u^A$. Because the reference
metric is stress free, the linear strain is simply $u_{ab}=\nabla_{(a}\delta u_{b)}$:
curvature introduces no additional terms. We take the lattice to move only within the surface, which is set by the trap. We split the displacement into compression and shear, $\delta u_a=\nabla_a\Phi+\epsilon_{ab}\nabla^b\Psi$, so that
the dilation is $\nabla^a\delta u_a=\Delta\Phi$. The equations of motion then read
\begin{subequations}
\label{eq:eom}
\begin{gather}
\chi\,\delta\ddot{\theta}
+g\,\Delta\dot{\Phi}
-\frac{\rho_s}{\mass^2}\Delta\delta\theta=0,
\\
\begin{aligned}
&\Delta\!\left[
  \rho_n\ddot{\Phi}
  +g\,\delta\dot{\theta}
  -(B+\mu_e)\Delta\Phi
\right]
-2\mu_e\nabla_a\!\left(K\nabla^a\Phi\right)\\
&\qquad
=2\mu_e\,\epsilon^{ab}
  \nabla_aK\,\nabla_b\Psi,
\end{aligned}
\\
\begin{aligned}
&\Delta\!\left[
  \rho_n\ddot{\Psi}
  -\mu_e\Delta\Psi
\right]
-2\mu_e\nabla_a\!\left(K\nabla^a\Psi\right)\\
&\qquad
=-2\mu_e\,\epsilon^{ab}
  \nabla_aK\,\nabla_b\Phi,
\end{aligned}
\end{gather}
\end{subequations}
where $K$ is the Gaussian curvature and $g=2\lambda-\rho_n/\mass$. Curvature enters
through the elastic force, which contains $\mu_eK\,\delta u_a$ because
$R_{ab}=Kg_{ab}$ in two dimensions. Equations~\eqref{eq:eom} hold
on any surface with the topology of a sphere and lead to three structural
conclusions. First, the phase couples only to compression, with strength $g$.
Second, compression and shear couple only through $\epsilon^{ab}\nabla_aK\nabla_b$,
which vanishes if and only if the curvature is uniform. Third, this coupling is
proportional to the shear modulus alone and involves no other material parameter:
it is a universal consequence of elasticity on a surface of non-uniform curvature.

Equations~\eqref{eq:eom} therefore define a chain of couplings,
shear\,$\xrightarrow{\nabla K}$\,compression\,$\xrightarrow{g}$\,superflow,
whose first link is broken on a sphere and restored by any deformation of it
(Fig.~\ref{fig:schematic}). As we show below, near collective-mode crossings this
chain is strong enough that a shear oscillation of the lattice carries as much
superfluid kinetic energy as lattice kinetic energy, however weak the deformation, which only sets the width of the crossing.

\begin{figure*}[ht!]
\centering
\includegraphics[width=\textwidth]{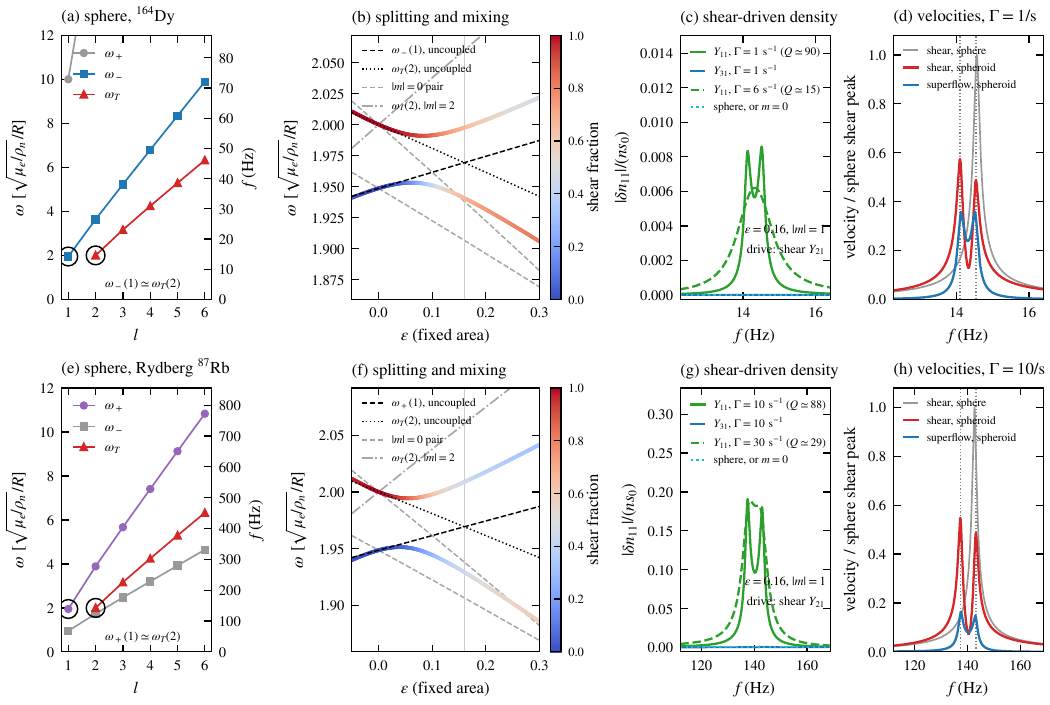}
\caption{Collective modes and shear-driven response for the two proposal materials: $^{164}$Dy in the top row (Table~\ref{tab:dy}), Rydberg-dressed $^{87}$Rb in the bottom row (Table~\ref{tab:ryd}).
(a,e)~Spectrum on the sphere, Eqs.~\eqref{eq:shear_sphere} and \eqref{eq:long_sphere}: longitudinal ($\omega_\pm$) and shear ($\omega_T$) branches; the $l=1$ shear modes are rigid rotations. Circles mark the near-degenerate pair. It is $\omega_-(1)\simeq\omega_T(2)$ in Dy, where $\omega_-$ is a counterflow mode, and $\omega_+(1)\simeq\omega_T(2)$ in the soft-core gas, where $\omega_+$ is mostly elastic. In both, $\omega_\sigma^2(1)$ lies 5\% below $\omega_T^2(2)$.
(b,f)~Splitting on a fixed-area spheroid, Eq.~\eqref{eq:splitting}, and mixing across $\Delta l=1$; the color gives the shear fraction. The $|m|=1$ levels shift with opposite slopes ($p_{11}p_{21}<0$) and anticross at $\varepsilon=0.16$ (vertical line), so both hybrids are nonmonotonic in $\varepsilon$. The $|m|=0$ pair (gray dashed) does not couple, and the $|m|=2$ shear level (dash-dotted) has no partner. Second-order corrections are of relative size $\varepsilon^2\le0.09$ over the plotted range.
(c,g)~Density response to a shear drive in $Y_{21}$, in units of $ns_0$, with $s_0=\Lambda_2|\hat\Psi|$ at the sphere resonance. It vanishes on the sphere and for $m=0$, and lies almost entirely in $Y_{11}$ ($Y_{31}<1\%$).
In Dy the peaks lie at 14.1 and 14.5~Hz with height $\simeq0.009\,ns_0$; they are resolved for $\Gamma=1$~s$^{-1}$ ($Q=\omega/\Gamma\simeq90$) and washed out for $\Gamma=6$~s$^{-1}$ ($Q\simeq15$).
In the soft-core gas they lie at 137.3 and 143.2~Hz with height $\simeq0.19\,ns_0$, about $20\times$ larger because the density is compressible; they are resolved for $\Gamma=10$~s$^{-1}$ ($Q\simeq88$) and only just resolved for $\Gamma=30$~s$^{-1}$ ($Q\simeq29$).
(d,h)~Velocities at the smaller $\Gamma$, normalized to the shear peak on the sphere.
In Dy, a counterflow--shear crossing, $|v_s|/|v_{\rm shear}|\simeq0.59$ and $0.69$ at the two peaks, close to $\mathcal R_\times\simeq0.73$ of Eq.~\eqref{eq:Rx}; between the hybrids the lattice shear drops to 0.13 while the superflow stays at 0.24.
In the soft-core gas, an elastic--shear crossing, $|v_s|/|v_{\rm shear}|\simeq0.29$, close to $\mathcal R_\times\simeq0.31$.
Parameters: Dy, $f_s=0.5$, $c_\theta^2/c_T^2=24$, $\lambda=0$, $B/\mu_e=3.96$, $c_T=0.69$~mm/s (frequency unit 7.3~Hz); soft-core, $f_s=0.5$, $c_\theta^2/c_T^2=0.6$, $g^2/(\chi\mu_e)=0.32$, $B/\mu_e=1.44$, $c_T=6.7$~mm/s (frequency unit 71~Hz); $R=15~\mu$m in both. All results are first order in $\varepsilon$.}
\label{fig:modes}
\end{figure*}

\textit{The sphere as a baseline.}---On a sphere of radius $R$ the curvature
$K=1/R^2$ is constant, the right-hand sides of Eqs.~\eqref{eq:eom} vanish, and every
field can be expanded in spherical harmonics $Y_{lm}$ with
$\Delta Y_{lm}=-\Lambda_lY_{lm}$, $\Lambda_l=l(l+1)/R^2$. The transverse (shear)
branch is
\begin{equation}
\rho_n\omega_T^2=\frac{\mu_e}{R^2}(l-1)(l+2),
\label{eq:shear_sphere}
\end{equation}
and carries neither phase nor density. Its $l=1$ modes have zero frequency because
they are the three rigid rotations of the sphere. The two longitudinal branches
$\omega_\pm$ are the roots of
\begin{equation}
\begin{aligned}
\left(\omega^2-\omega_\theta^2\right)
\left(\omega^2-\omega_E^2\right)
&=\frac{g^2\Lambda_l}{\chi\rho_n}\,\omega^2,\\
\omega_\theta^2
&=\frac{\rho_s\Lambda_l}{\mass^2\chi},\\
\omega_E^2
&=\frac{(B+\mu_e)\Lambda_l-2\mu_e/R^2}{\rho_n},
\end{aligned}
\label{eq:long_sphere}
\end{equation}
i.e., the phase mode of the condensate and the compressional mode of the lattice,
hybridized by $g$ [Fig.~\ref{fig:modes}(a)]. Curvature softens both elastic branches
by $2\mu_e/R^2$, and for $l\gg1$ the three branches reduce to the transverse and the
two longitudinal sounds of a flat supersolid~\cite{yoo_hydrodynamic_2010}. We denote
by $Z_\pm$ the elastic weight of each longitudinal branch (the fraction of the mode
carried by $\Phi$; see SM). Each level is $(2l+1)$-fold degenerate in $m$.

\textit{Spheroid: splitting of the multiplets.}---We now deform the sphere into a
spheroid with semi-axes $a$ (equatorial) and $c$ (polar), $c^2=a^2(1+\varepsilon)$,
keeping the area fixed at $4\pi R^2$. By the Gauss--Bonnet theorem the area-averaged Gaussian curvature then remains $1/R^2$, and the two surfaces differ only by the quadrupolar
modulation $K\simeq R^{-2}[1+\varepsilon(\cos2\vartheta+\tfrac13)]$. Rotations about the
axis remain a symmetry, so $m$ is conserved.
To first order in $\varepsilon$ every branch $s\in\{T,+,-\}$ splits as
\begin{equation}
\omega_s^2(\varepsilon)
=\omega_s^2\left[1-\varepsilon\,\kappa_s(l)\,p_{lm}\right],
\label{eq:splitting}
\end{equation}
where \(p_{lm}=[l(l+1)-3m^2]/[(2l-1)(2l+3)]\) and
\(\kappa_s(l)=2/3+4Z_s\omega_T^2(l)/[l(l+1)\omega_s^2]\), with
\(Z_T=1\) and all quantities on the right evaluated on the sphere. The
$m$-dependence enters only through the quadrupole weight $p_{lm}$; it is fixed by
geometry and identical for all branches, analogous to the ellipticity-induced splitting of Earth's free-oscillation multiplets \cite{Dahlen1968}. The material enters only through the slope $\kappa_s$:
$2/3$ is the shift of the Laplacian spectrum, common to all branches, and the second
term comes from the curvature term of the elastic force, weighted by the elastic
content of the branch. For the shear branch $\kappa_T=2/3+4/[l(l+1)]$ is
material independent, so the shear multiplets measure the deformation alone.

\textit{Selected crossings and their mixing.}---Beyond splitting the multiplets, the
deformation couples harmonics with different $l$ but equal $m$. The parity of the
spheroid under $z\to-z$ yields selection rules. The term
$\epsilon^{ab}\nabla_aK\nabla_b\propto\partial_\vartheta K\,\partial_\phi$ is odd and
proportional to $m$: it couples compression in $Y_{lm}$ to shear in $Y_{l\pm1,m}$,
and only for $m\neq0$. The even parts of the deformation couple each field to itself
across $\Delta l=\pm2$, and $|\Delta l|\geq3$ is not coupled at first order. Levels of the same type never cross, and the physically relevant degeneracies are those
between a longitudinal level of $l$ and the shear level of $l+1$. Near such a crossing the two levels repel, with minimal splitting
\begin{equation}
\Delta(\omega^2)_{\min}
=\frac{16|\varepsilon|\mu_e}{\rho_nR^2}\,
\frac{|m|\,J_{lm}}{\sqrt{l(l+1)(l+1)(l+2)}}\,\sqrt{Z_\sigma},
\label{eq:gap}
\end{equation}
where $J_{lm}=\sqrt{\frac{(l+1)^2-m^2}{(2l+1)(2l+3)}}.$ The gap is proportional to $\varepsilon m$ and to $\sqrt{Z_\sigma}$, because the
shear mode only interacts with the compressional content of the longitudinal level.
Hence only specific pairs hybridize: equal $m$, $\Delta l=\pm1$, $m\neq0$, and a
longitudinal partner with an elastic component. Figure~\ref{fig:modes}(b) shows this for the longitudinal level $\omega_-(1)$ and the shear level $\omega_T(2)$. All multiplets fan out according to Eq.~\eqref{eq:splitting}; the $|m|=1$ pair anticrosses, the $|m|=0$ pair crosses freely, and the shear $|m|=2$ level has no partner. A further class of
crossings, $\omega_+(l)=\omega_-(l+2)$, pairs the phase mode of one multiplet with
the elastic mode of another; it exists only in a supersolid, survives for $m=0$, and
its gap requires both $g$ and $\nabla K$; we have colored the modes according to the $\Delta l = \pm1$ crossing (SM). 

Because the crossings can also be reached by changing the aspect ratio, curvature gradients leave a directly measurable fingerprint. Along $\varepsilon$ each level moves linearly, and near a crossing the repulsion exchanges the two lines between the roots. When the two levels shift with opposite slopes, $p_{lm}p_{l'm}<0$, i.e.\ $l(l+1)<3m^2<l'(l'+1)$ [e.g.\ $(l,|m|)=(1,1),(4,3),(6,4)$], the frequency of each hybrid therefore becomes a nonmonotonic function of $\varepsilon$, with a maximum for the lower root and a minimum for the upper one (see SM). This condition is independent of the material, and it never holds for $m=0$, where the slopes have equal signs and the levels do not couple. A turning point of a mode frequency versus aspect ratio thus signals the compression--shear coupling $\epsilon^{ab}\nabla_aK\nabla_b$.


\textit{Shear-driven superflow.}---We now drive the lattice with a pure shear force,
$f_a=\epsilon_{ab}\nabla^b\psi_f$ with $\psi_f=\hat\psi_fY_{Lm}e^{-i\omega t}$. It has
no divergence and on a sphere excites only $\Psi$. On the spheroid, the selection
rules feed compression and phase in the sectors $k=L\pm1$, and the shear response
becomes
\begin{equation}
\begin{aligned}
\hat\Psi
&=-\frac{\hat\psi_f}{\rho_n}\,
  \frac{1}{\omega^2-\omega_T^2-\Sigma(\omega^2)},\\
\Sigma(x)
&=\sum_{k=L\pm1}\sum_{\sigma=\pm}
  \frac{\hat V_k^{\,2}Z_{\sigma,k}}
  {\rho_n^2\bigl(x-\omega_{\sigma,k}^2\bigr)}
  +\cdots ,
\end{aligned}
\label{eq:response}
\end{equation}
where $\hat V_k\propto\varepsilon mJ$ is the $\Delta l=\pm1$ coupling and the dots
denote the weaker $\Delta l=\pm2$ shear-shear terms (SM). The poles of $\Sigma$ are the
levels coupled to the drive, and its residues are the squared mixing amplitudes. The
superfluid velocity and the density in sector $k$ share the same denominator as
$\hat\Psi$, and three properties follow. First, the superfluid responds only at
frequencies of modes that contain shear: the uncoupled longitudinal poles cancel.
Second, superflow and density are proportional to $\hat V_k$; they vanish on the
sphere and for $m=0$, where the $\Delta l=\pm2$ mixing alone only redistributes
weight among shear modes. Third, the frequency window of strong response is set by
the gap~\eqref{eq:gap}, of width $\propto\varepsilon m\sqrt{Z_\sigma}$
[Fig.~\ref{fig:modes}(c,g)].

The height of the response follows from the content of the hybrid modes. Away from
crossings the ratio of superflow to shear velocity is $O(\varepsilon mg)$. At a
crossing $\omega_T(L)=\omega_\sigma(k)$, the coupling cancels and
\begin{equation}
\mathcal{R}_\times\equiv\frac{|v_s|}{|v_{\rm shear}|}
=\frac{1}{\mass\,\omega_\sigma}\sqrt{\frac{\rho_n\Lambda_kY_\sigma}{\chi}},
\label{eq:Rx}
\end{equation}
with $Y_\sigma$ the phase weight of the longitudinal level. $\mathcal{R}_\times$ is
independent of $\varepsilon$ and of the azimuthal number $m$: the geometry determines where the crossing is
and how wide it is, while the material determines how much superflow the hybrid
carries. At a phase--shear crossing with weak $g$, $\mathcal{R}_\times^2\simeq
\rho_n/\rho_s$, i.e. $\rho_s|v_s|^2\simeq\rho_n|v_{\rm shear}|^2$: the kinetic energy
of the hybrid mode is shared equally between the superflow and the lattice shear,
however small the deformation. At an elastic--shear crossing, by contrast,
$\mathcal{R}_\times=O(g)$.

\textit{Experimental signatures.}---Through the $\Delta l=\pm1$ mixing, a shear drive in $Y_{Lm}$ produces compression and phase, and hence a density modulation $\delta\hat n_k=-i\omega\chi\hat\theta_k-2\lambda\Lambda_k\hat\Phi_k$, where $\hat\theta_k, \hat\Phi_k$ are the phase and compression amplitudes in $Y_{k,m}$. On the sphere, or for $m=0$, this response vanishes identically. A shear excitation that density probes cannot see on a sphere thus becomes visible on a spheroid, as a doublet at the two hybrids, and comparing the two shapes reveals the curvature-gradient coupling without phase imaging. Our theory is non-dissipative, so the response has poles at the hybrids. To compare with experiments we add a small phenomenological damping $\omega^2 \to \omega^2 + i \Gamma\omega$, with values set by lifetime (SM), and it is resolved when the value is smaller than the splitting of Eq.~\eqref{eq:gap}.

Such a comparison needs a supersolid on a shell. Our results depend on the material only through four dimensionless ratios of the coefficients, and give frequencies in units of $c_T/R$, with $c_T=\sqrt{\mu_e/\rho_n}$ the transverse sound speed. In flat space the longitudinal modes become the first and second sounds of the supersolid, $c_+>c_-$, the two hybrids of the phase mode and the lattice compression. Microscopic calculations give these speeds and the elastic parameters rather than our coefficients, so we translate them, and convert frequencies with $f=\omega\,c_T/(2\pi R)$ (SM). Figure~\ref{fig:modes} compares two candidate platforms, which probe the two kinds of crossing. A $^{164}$Dy dipolar supersolid~\cite{Norcia2021TwoDimensional} realizes the counterflow--shear crossing, where the hybrids carry nearly as much superfluid as lattice kinetic energy; its density signal is weak, since only the phase moves the density, and resolving the doublet requires second-long lifetimes~\cite{Sohmen2021Birth}. A Rydberg-dressed gas, whose isotropic soft-core interaction can form a cluster crystal over the whole shell~\cite{Cinti2010Droplet,Ciardi2024BubbleSupersolids}, realizes the elastic--shear crossing instead: compression adds to the density, giving a signal an order of magnitude larger that survives the damping expected from dressed lifetimes~\cite{Weckesser2025Dressed}, at the price of a weaker superflow. Such a supersolid has not yet been observed.

\textit{Conclusion.}---We have developed an effective field theory of supersolids on curved surfaces and uncovered a universal geometric mechanism by which curvature gradients couple lattice shear to compression and superfluid flow. The shear–-compression coupling is fixed solely by the shear modulus and $\nabla K$, independent of microscopic details. Geometry thus controls collective-mode hybridization through symmetry-protected selection rules, enabling even weak surface deformations to generate strong shear–-superflow mixing near mode crossings.

{\it{Acknowledgments}}$-$We thank Abel Beregi, Jan Michael Rost, Andrea Tononi, Lazaros Tsaloukidis, and Marek Tylutki for useful discussions. P.S. was supported in part by the Polish National Science Centre (NCN) Sonata Bis grant 2019/34/E/ST3/00405. Y.O.S. was supported in part by IMPRS-MPSSE. The code used to generate the figures was written with the assistance of Claude (Anthropic).
 
\bibliography{references}

\begin{thebibliography}{41}%
\makeatletter
\providecommand \@ifxundefined [1]{%
 \@ifx{#1\undefined}
}%
\providecommand \@ifnum [1]{%
 \ifnum #1\expandafter \@firstoftwo
 \else \expandafter \@secondoftwo
 \fi
}%
\providecommand \@ifx [1]{%
 \ifx #1\expandafter \@firstoftwo
 \else \expandafter \@secondoftwo
 \fi
}%
\providecommand \natexlab [1]{#1}%
\providecommand \enquote  [1]{``#1''}%
\providecommand \bibnamefont  [1]{#1}%
\providecommand \bibfnamefont [1]{#1}%
\providecommand \citenamefont [1]{#1}%
\providecommand \href@noop [0]{\@secondoftwo}%
\providecommand \href [0]{\begingroup \@sanitize@url \@href}%
\providecommand \@href[1]{\@@startlink{#1}\@@href}%
\providecommand \@@href[1]{\endgroup#1\@@endlink}%
\providecommand \@sanitize@url [0]{\catcode `\\12\catcode `\$12\catcode `\&12\catcode `\#12\catcode `\^12\catcode `\_12\catcode `\%12\relax}%
\providecommand \@@startlink[1]{}%
\providecommand \@@endlink[0]{}%
\providecommand \url  [0]{\begingroup\@sanitize@url \@url }%
\providecommand \@url [1]{\endgroup\@href {#1}{\urlprefix }}%
\providecommand \urlprefix  [0]{URL }%
\providecommand \Eprint [0]{\href }%
\providecommand \doibase [0]{https://doi.org/}%
\providecommand \selectlanguage [0]{\@gobble}%
\providecommand \bibinfo  [0]{\@secondoftwo}%
\providecommand \bibfield  [0]{\@secondoftwo}%
\providecommand \translation [1]{[#1]}%
\providecommand \BibitemOpen [0]{}%
\providecommand \bibitemStop [0]{}%
\providecommand \bibitemNoStop [0]{.\EOS\space}%
\providecommand \EOS [0]{\spacefactor3000\relax}%
\providecommand \BibitemShut  [1]{\csname bibitem#1\endcsname}%
\let\auto@bib@innerbib\@empty
\bibitem [{\citenamefont {Andreev}\ and\ \citenamefont {Lifshitz}(1969)}]{Andreev1969}%
  \BibitemOpen
  \bibfield  {author} {\bibinfo {author} {\bibfnamefont {A.~F.}\ \bibnamefont {Andreev}}\ and\ \bibinfo {author} {\bibfnamefont {I.~M.}\ \bibnamefont {Lifshitz}},\ }\bibfield  {title} {\bibinfo {title} {Quantum theory of defects in crystals},\ }\href@noop {} {\bibfield  {journal} {\bibinfo  {journal} {Soviet Physics JETP}\ }\textbf {\bibinfo {volume} {29}},\ \bibinfo {pages} {1107} (\bibinfo {year} {1969})}\BibitemShut {NoStop}%
\bibitem [{\citenamefont {Chester}(1970)}]{Chester1970}%
  \BibitemOpen
  \bibfield  {author} {\bibinfo {author} {\bibfnamefont {G.~V.}\ \bibnamefont {Chester}},\ }\bibfield  {title} {\bibinfo {title} {Speculations on {Bose--Einstein} condensation and quantum crystals},\ }\href {https://doi.org/10.1103/PhysRevA.2.256} {\bibfield  {journal} {\bibinfo  {journal} {Physical Review A}\ }\textbf {\bibinfo {volume} {2}},\ \bibinfo {pages} {256} (\bibinfo {year} {1970})}\BibitemShut {NoStop}%
\bibitem [{\citenamefont {Leggett}(1970)}]{Leggett1970}%
  \BibitemOpen
  \bibfield  {author} {\bibinfo {author} {\bibfnamefont {A.~J.}\ \bibnamefont {Leggett}},\ }\bibfield  {title} {\bibinfo {title} {Can a solid be ``superfluid''?},\ }\href {https://doi.org/10.1103/PhysRevLett.25.1543} {\bibfield  {journal} {\bibinfo  {journal} {Physical Review Letters}\ }\textbf {\bibinfo {volume} {25}},\ \bibinfo {pages} {1543} (\bibinfo {year} {1970})}\BibitemShut {NoStop}%
\bibitem [{\citenamefont {Recati}\ and\ \citenamefont {Stringari}(2023)}]{Recati2023Supersolidity}%
  \BibitemOpen
  \bibfield  {author} {\bibinfo {author} {\bibfnamefont {A.}~\bibnamefont {Recati}}\ and\ \bibinfo {author} {\bibfnamefont {S.}~\bibnamefont {Stringari}},\ }\bibfield  {title} {\bibinfo {title} {Supersolidity in ultracold dipolar gases},\ }\href {https://doi.org/10.1038/s42254-023-00648-2} {\bibfield  {journal} {\bibinfo  {journal} {Nature Reviews Physics}\ }\textbf {\bibinfo {volume} {5}},\ \bibinfo {pages} {735} (\bibinfo {year} {2023})}\BibitemShut {NoStop}%
\bibitem [{\citenamefont {Li}\ \emph {et~al.}(2017)\citenamefont {Li}, \citenamefont {Lee}, \citenamefont {Huang}, \citenamefont {Burchesky}, \citenamefont {Shteynas}, \citenamefont {Top}, \citenamefont {Jamison},\ and\ \citenamefont {Ketterle}}]{Li2017Stripe}%
  \BibitemOpen
  \bibfield  {author} {\bibinfo {author} {\bibfnamefont {J.-R.}\ \bibnamefont {Li}}, \bibinfo {author} {\bibfnamefont {J.}~\bibnamefont {Lee}}, \bibinfo {author} {\bibfnamefont {W.}~\bibnamefont {Huang}}, \bibinfo {author} {\bibfnamefont {S.}~\bibnamefont {Burchesky}}, \bibinfo {author} {\bibfnamefont {B.}~\bibnamefont {Shteynas}}, \bibinfo {author} {\bibfnamefont {F.~{\c{C}}.}\ \bibnamefont {Top}}, \bibinfo {author} {\bibfnamefont {A.~O.}\ \bibnamefont {Jamison}},\ and\ \bibinfo {author} {\bibfnamefont {W.}~\bibnamefont {Ketterle}},\ }\bibfield  {title} {\bibinfo {title} {A stripe phase with supersolid properties in spin--orbit-coupled {Bose--Einstein} condensates},\ }\href {https://doi.org/10.1038/nature21431} {\bibfield  {journal} {\bibinfo  {journal} {Nature}\ }\textbf {\bibinfo {volume} {543}},\ \bibinfo {pages} {91} (\bibinfo {year} {2017})}\BibitemShut {NoStop}%
\bibitem [{\citenamefont {L{\'e}onard}\ \emph {et~al.}(2017)\citenamefont {L{\'e}onard}, \citenamefont {Morales}, \citenamefont {Zupancic}, \citenamefont {Esslinger},\ and\ \citenamefont {Donner}}]{Leonard2017Supersolid}%
  \BibitemOpen
  \bibfield  {author} {\bibinfo {author} {\bibfnamefont {J.}~\bibnamefont {L{\'e}onard}}, \bibinfo {author} {\bibfnamefont {A.}~\bibnamefont {Morales}}, \bibinfo {author} {\bibfnamefont {P.}~\bibnamefont {Zupancic}}, \bibinfo {author} {\bibfnamefont {T.}~\bibnamefont {Esslinger}},\ and\ \bibinfo {author} {\bibfnamefont {T.}~\bibnamefont {Donner}},\ }\bibfield  {title} {\bibinfo {title} {Supersolid formation in a quantum gas breaking a continuous translational symmetry},\ }\href {https://doi.org/10.1038/nature21067} {\bibfield  {journal} {\bibinfo  {journal} {Nature}\ }\textbf {\bibinfo {volume} {543}},\ \bibinfo {pages} {87} (\bibinfo {year} {2017})}\BibitemShut {NoStop}%
\bibitem [{\citenamefont {B{\"o}ttcher}\ \emph {et~al.}(2019)\citenamefont {B{\"o}ttcher}, \citenamefont {Schmidt}, \citenamefont {Wenzel}, \citenamefont {Hertkorn}, \citenamefont {Guo}, \citenamefont {Langen},\ and\ \citenamefont {Pfau}}]{Bottcher2019Transient}%
  \BibitemOpen
  \bibfield  {author} {\bibinfo {author} {\bibfnamefont {F.}~\bibnamefont {B{\"o}ttcher}}, \bibinfo {author} {\bibfnamefont {J.-N.}\ \bibnamefont {Schmidt}}, \bibinfo {author} {\bibfnamefont {M.}~\bibnamefont {Wenzel}}, \bibinfo {author} {\bibfnamefont {J.}~\bibnamefont {Hertkorn}}, \bibinfo {author} {\bibfnamefont {M.}~\bibnamefont {Guo}}, \bibinfo {author} {\bibfnamefont {T.}~\bibnamefont {Langen}},\ and\ \bibinfo {author} {\bibfnamefont {T.}~\bibnamefont {Pfau}},\ }\bibfield  {title} {\bibinfo {title} {Transient supersolid properties in an array of dipolar quantum droplets},\ }\href {https://doi.org/10.1103/PhysRevX.9.011051} {\bibfield  {journal} {\bibinfo  {journal} {Physical Review X}\ }\textbf {\bibinfo {volume} {9}},\ \bibinfo {pages} {011051} (\bibinfo {year} {2019})}\BibitemShut {NoStop}%
\bibitem [{\citenamefont {Tanzi}\ \emph {et~al.}(2019{\natexlab{a}})\citenamefont {Tanzi}, \citenamefont {Lucioni}, \citenamefont {Fam{\`a}}, \citenamefont {Catani}, \citenamefont {Fioretti}, \citenamefont {Gabbanini}, \citenamefont {Bisset}, \citenamefont {Santos},\ and\ \citenamefont {Modugno}}]{Tanzi2019Observation}%
  \BibitemOpen
  \bibfield  {author} {\bibinfo {author} {\bibfnamefont {L.}~\bibnamefont {Tanzi}}, \bibinfo {author} {\bibfnamefont {E.}~\bibnamefont {Lucioni}}, \bibinfo {author} {\bibfnamefont {F.}~\bibnamefont {Fam{\`a}}}, \bibinfo {author} {\bibfnamefont {J.}~\bibnamefont {Catani}}, \bibinfo {author} {\bibfnamefont {A.}~\bibnamefont {Fioretti}}, \bibinfo {author} {\bibfnamefont {C.}~\bibnamefont {Gabbanini}}, \bibinfo {author} {\bibfnamefont {R.~N.}\ \bibnamefont {Bisset}}, \bibinfo {author} {\bibfnamefont {L.}~\bibnamefont {Santos}},\ and\ \bibinfo {author} {\bibfnamefont {G.}~\bibnamefont {Modugno}},\ }\bibfield  {title} {\bibinfo {title} {Observation of a dipolar quantum gas with metastable supersolid properties},\ }\href {https://doi.org/10.1103/PhysRevLett.122.130405} {\bibfield  {journal} {\bibinfo  {journal} {Physical Review Letters}\ }\textbf {\bibinfo {volume} {122}},\ \bibinfo {pages} {130405} (\bibinfo {year} {2019}{\natexlab{a}})}\BibitemShut {NoStop}%
\bibitem [{\citenamefont {Chomaz}\ \emph {et~al.}(2019)\citenamefont {Chomaz}, \citenamefont {Petter}, \citenamefont {Ilzh{\"o}fer}, \citenamefont {Natale}, \citenamefont {Trautmann}, \citenamefont {Politi}, \citenamefont {Durastante}, \citenamefont {van Bijnen}, \citenamefont {Patscheider}, \citenamefont {Sohmen}, \citenamefont {Mark},\ and\ \citenamefont {Ferlaino}}]{Chomaz2019LongLived}%
  \BibitemOpen
  \bibfield  {author} {\bibinfo {author} {\bibfnamefont {L.}~\bibnamefont {Chomaz}}, \bibinfo {author} {\bibfnamefont {D.}~\bibnamefont {Petter}}, \bibinfo {author} {\bibfnamefont {P.}~\bibnamefont {Ilzh{\"o}fer}}, \bibinfo {author} {\bibfnamefont {G.}~\bibnamefont {Natale}}, \bibinfo {author} {\bibfnamefont {A.}~\bibnamefont {Trautmann}}, \bibinfo {author} {\bibfnamefont {C.}~\bibnamefont {Politi}}, \bibinfo {author} {\bibfnamefont {G.}~\bibnamefont {Durastante}}, \bibinfo {author} {\bibfnamefont {R.~M.~W.}\ \bibnamefont {van Bijnen}}, \bibinfo {author} {\bibfnamefont {A.}~\bibnamefont {Patscheider}}, \bibinfo {author} {\bibfnamefont {M.}~\bibnamefont {Sohmen}}, \bibinfo {author} {\bibfnamefont {M.~J.}\ \bibnamefont {Mark}},\ and\ \bibinfo {author} {\bibfnamefont {F.}~\bibnamefont {Ferlaino}},\ }\bibfield  {title} {\bibinfo {title} {Long-lived and transient supersolid behaviors in dipolar quantum gases},\ }\href {https://doi.org/10.1103/PhysRevX.9.021012} {\bibfield  {journal} {\bibinfo  {journal} {Physical
  Review X}\ }\textbf {\bibinfo {volume} {9}},\ \bibinfo {pages} {021012} (\bibinfo {year} {2019})}\BibitemShut {NoStop}%
\bibitem [{\citenamefont {Norcia}\ \emph {et~al.}(2021)\citenamefont {Norcia}, \citenamefont {Politi}, \citenamefont {Klaus}, \citenamefont {Poli}, \citenamefont {Sohmen}, \citenamefont {Mark}, \citenamefont {Bisset}, \citenamefont {Santos},\ and\ \citenamefont {Ferlaino}}]{Norcia2021TwoDimensional}%
  \BibitemOpen
  \bibfield  {author} {\bibinfo {author} {\bibfnamefont {M.~A.}\ \bibnamefont {Norcia}}, \bibinfo {author} {\bibfnamefont {C.}~\bibnamefont {Politi}}, \bibinfo {author} {\bibfnamefont {L.}~\bibnamefont {Klaus}}, \bibinfo {author} {\bibfnamefont {E.}~\bibnamefont {Poli}}, \bibinfo {author} {\bibfnamefont {M.}~\bibnamefont {Sohmen}}, \bibinfo {author} {\bibfnamefont {M.~J.}\ \bibnamefont {Mark}}, \bibinfo {author} {\bibfnamefont {R.~N.}\ \bibnamefont {Bisset}}, \bibinfo {author} {\bibfnamefont {L.}~\bibnamefont {Santos}},\ and\ \bibinfo {author} {\bibfnamefont {F.}~\bibnamefont {Ferlaino}},\ }\bibfield  {title} {\bibinfo {title} {Two-dimensional supersolidity in a dipolar quantum gas},\ }\href {https://doi.org/10.1038/s41586-021-03725-7} {\bibfield  {journal} {\bibinfo  {journal} {Nature}\ }\textbf {\bibinfo {volume} {596}},\ \bibinfo {pages} {357} (\bibinfo {year} {2021})}\BibitemShut {NoStop}%
\bibitem [{\citenamefont {Tanzi}\ \emph {et~al.}(2019{\natexlab{b}})\citenamefont {Tanzi}, \citenamefont {Roccuzzo}, \citenamefont {Lucioni}, \citenamefont {Fam{\`a}}, \citenamefont {Fioretti}, \citenamefont {Gabbanini}, \citenamefont {Modugno}, \citenamefont {Recati},\ and\ \citenamefont {Stringari}}]{Tanzi2019SymmetryBreaking}%
  \BibitemOpen
  \bibfield  {author} {\bibinfo {author} {\bibfnamefont {L.}~\bibnamefont {Tanzi}}, \bibinfo {author} {\bibfnamefont {S.~M.}\ \bibnamefont {Roccuzzo}}, \bibinfo {author} {\bibfnamefont {E.}~\bibnamefont {Lucioni}}, \bibinfo {author} {\bibfnamefont {F.}~\bibnamefont {Fam{\`a}}}, \bibinfo {author} {\bibfnamefont {A.}~\bibnamefont {Fioretti}}, \bibinfo {author} {\bibfnamefont {C.}~\bibnamefont {Gabbanini}}, \bibinfo {author} {\bibfnamefont {G.}~\bibnamefont {Modugno}}, \bibinfo {author} {\bibfnamefont {A.}~\bibnamefont {Recati}},\ and\ \bibinfo {author} {\bibfnamefont {S.}~\bibnamefont {Stringari}},\ }\bibfield  {title} {\bibinfo {title} {Supersolid symmetry breaking from compressional oscillations in a dipolar quantum gas},\ }\href {https://doi.org/10.1038/s41586-019-1568-6} {\bibfield  {journal} {\bibinfo  {journal} {Nature}\ }\textbf {\bibinfo {volume} {574}},\ \bibinfo {pages} {382} (\bibinfo {year} {2019}{\natexlab{b}})}\BibitemShut {NoStop}%
\bibitem [{\citenamefont {Guo}\ \emph {et~al.}(2019)\citenamefont {Guo}, \citenamefont {B{\"o}ttcher}, \citenamefont {Hertkorn}, \citenamefont {Schmidt}, \citenamefont {Wenzel}, \citenamefont {B{\"u}chler}, \citenamefont {Langen},\ and\ \citenamefont {Pfau}}]{Guo2019Goldstone}%
  \BibitemOpen
  \bibfield  {author} {\bibinfo {author} {\bibfnamefont {M.}~\bibnamefont {Guo}}, \bibinfo {author} {\bibfnamefont {F.}~\bibnamefont {B{\"o}ttcher}}, \bibinfo {author} {\bibfnamefont {J.}~\bibnamefont {Hertkorn}}, \bibinfo {author} {\bibfnamefont {J.-N.}\ \bibnamefont {Schmidt}}, \bibinfo {author} {\bibfnamefont {M.}~\bibnamefont {Wenzel}}, \bibinfo {author} {\bibfnamefont {H.~P.}\ \bibnamefont {B{\"u}chler}}, \bibinfo {author} {\bibfnamefont {T.}~\bibnamefont {Langen}},\ and\ \bibinfo {author} {\bibfnamefont {T.}~\bibnamefont {Pfau}},\ }\bibfield  {title} {\bibinfo {title} {The low-energy {Goldstone} mode in a trapped dipolar supersolid},\ }\href {https://doi.org/10.1038/s41586-019-1569-5} {\bibfield  {journal} {\bibinfo  {journal} {Nature}\ }\textbf {\bibinfo {volume} {574}},\ \bibinfo {pages} {386} (\bibinfo {year} {2019})}\BibitemShut {NoStop}%
\bibitem [{\citenamefont {Casotti}\ \emph {et~al.}(2024)\citenamefont {Casotti}, \citenamefont {Poli}, \citenamefont {Klaus}, \citenamefont {Litvinov}, \citenamefont {Ulm}, \citenamefont {Politi}, \citenamefont {Mark}, \citenamefont {Bland},\ and\ \citenamefont {Ferlaino}}]{Casotti2024Vortices}%
  \BibitemOpen
  \bibfield  {author} {\bibinfo {author} {\bibfnamefont {E.}~\bibnamefont {Casotti}}, \bibinfo {author} {\bibfnamefont {E.}~\bibnamefont {Poli}}, \bibinfo {author} {\bibfnamefont {L.}~\bibnamefont {Klaus}}, \bibinfo {author} {\bibfnamefont {A.}~\bibnamefont {Litvinov}}, \bibinfo {author} {\bibfnamefont {C.}~\bibnamefont {Ulm}}, \bibinfo {author} {\bibfnamefont {C.}~\bibnamefont {Politi}}, \bibinfo {author} {\bibfnamefont {M.~J.}\ \bibnamefont {Mark}}, \bibinfo {author} {\bibfnamefont {T.}~\bibnamefont {Bland}},\ and\ \bibinfo {author} {\bibfnamefont {F.}~\bibnamefont {Ferlaino}},\ }\bibfield  {title} {\bibinfo {title} {Observation of vortices in a dipolar supersolid},\ }\href {https://doi.org/10.1038/s41586-024-08149-7} {\bibfield  {journal} {\bibinfo  {journal} {Nature}\ }\textbf {\bibinfo {volume} {635}},\ \bibinfo {pages} {327} (\bibinfo {year} {2024})}\BibitemShut {NoStop}%
\bibitem [{\citenamefont {Acosta~Matos}\ \emph {et~al.}(2025)\citenamefont {Acosta~Matos}, \citenamefont {Giannakeas}, \citenamefont {Ciardi}, \citenamefont {Pohl},\ and\ \citenamefont {Rost}}]{AcostaMatos2025Transitional}%
  \BibitemOpen
  \bibfield  {author} {\bibinfo {author} {\bibfnamefont {J.~C.}\ \bibnamefont {Acosta~Matos}}, \bibinfo {author} {\bibfnamefont {P.}~\bibnamefont {Giannakeas}}, \bibinfo {author} {\bibfnamefont {M.}~\bibnamefont {Ciardi}}, \bibinfo {author} {\bibfnamefont {T.}~\bibnamefont {Pohl}},\ and\ \bibinfo {author} {\bibfnamefont {J.~M.}\ \bibnamefont {Rost}},\ }\bibfield  {title} {\bibinfo {title} {Transitional supersolidity in ion-doped helium droplets},\ }\href {https://doi.org/10.1103/PhysRevLett.134.186001} {\bibfield  {journal} {\bibinfo  {journal} {Physical Review Letters}\ }\textbf {\bibinfo {volume} {134}},\ \bibinfo {pages} {186001} (\bibinfo {year} {2025})}\BibitemShut {NoStop}%
\bibitem [{\citenamefont {Tononi}\ and\ \citenamefont {Salasnich}(2023)}]{Tononi2023CurvedGases}%
  \BibitemOpen
  \bibfield  {author} {\bibinfo {author} {\bibfnamefont {A.}~\bibnamefont {Tononi}}\ and\ \bibinfo {author} {\bibfnamefont {L.}~\bibnamefont {Salasnich}},\ }\bibfield  {title} {\bibinfo {title} {Low-dimensional quantum gases in curved geometries},\ }\href {https://doi.org/10.1038/s42254-023-00591-2} {\bibfield  {journal} {\bibinfo  {journal} {Nature Reviews Physics}\ }\textbf {\bibinfo {volume} {5}},\ \bibinfo {pages} {398} (\bibinfo {year} {2023})}\BibitemShut {NoStop}%
\bibitem [{\citenamefont {Tononi}\ and\ \citenamefont {Astrakharchik}(2025)}]{Tononi2025AxiallySymmetric}%
  \BibitemOpen
  \bibfield  {author} {\bibinfo {author} {\bibfnamefont {A.}~\bibnamefont {Tononi}}\ and\ \bibinfo {author} {\bibfnamefont {G.~E.}\ \bibnamefont {Astrakharchik}},\ }\bibfield  {title} {\bibinfo {title} {{Bose--Einstein} condensation on axially symmetric surfaces},\ }\href {https://doi.org/10.1103/vxy6-skdq} {\bibfield  {journal} {\bibinfo  {journal} {Physical Review A}\ }\textbf {\bibinfo {volume} {112}},\ \bibinfo {pages} {023302} (\bibinfo {year} {2025})}\BibitemShut {NoStop}%
\bibitem [{\citenamefont {Carollo}\ \emph {et~al.}(2022)\citenamefont {Carollo}, \citenamefont {Aveline}, \citenamefont {Rhyno}, \citenamefont {Vishveshwara}, \citenamefont {Lannert}, \citenamefont {Murphree}, \citenamefont {Elliott}, \citenamefont {Williams}, \citenamefont {Thompson},\ and\ \citenamefont {Lundblad}}]{Carollo2022AtomicBubbles}%
  \BibitemOpen
  \bibfield  {author} {\bibinfo {author} {\bibfnamefont {R.~A.}\ \bibnamefont {Carollo}}, \bibinfo {author} {\bibfnamefont {D.~C.}\ \bibnamefont {Aveline}}, \bibinfo {author} {\bibfnamefont {B.}~\bibnamefont {Rhyno}}, \bibinfo {author} {\bibfnamefont {S.}~\bibnamefont {Vishveshwara}}, \bibinfo {author} {\bibfnamefont {C.}~\bibnamefont {Lannert}}, \bibinfo {author} {\bibfnamefont {J.~D.}\ \bibnamefont {Murphree}}, \bibinfo {author} {\bibfnamefont {E.~R.}\ \bibnamefont {Elliott}}, \bibinfo {author} {\bibfnamefont {J.~R.}\ \bibnamefont {Williams}}, \bibinfo {author} {\bibfnamefont {R.~J.}\ \bibnamefont {Thompson}},\ and\ \bibinfo {author} {\bibfnamefont {N.}~\bibnamefont {Lundblad}},\ }\bibfield  {title} {\bibinfo {title} {Observation of ultracold atomic bubbles in orbital microgravity},\ }\href {https://doi.org/10.1038/s41586-022-04639-8} {\bibfield  {journal} {\bibinfo  {journal} {Nature}\ }\textbf {\bibinfo {volume} {606}},\ \bibinfo {pages} {281} (\bibinfo {year} {2022})}\BibitemShut {NoStop}%
\bibitem [{\citenamefont {Nicolis}\ \emph {et~al.}(2015)\citenamefont {Nicolis}, \citenamefont {Penco}, \citenamefont {Piazza},\ and\ \citenamefont {Rattazzi}}]{Nicolis2015Zoology}%
  \BibitemOpen
  \bibfield  {author} {\bibinfo {author} {\bibfnamefont {A.}~\bibnamefont {Nicolis}}, \bibinfo {author} {\bibfnamefont {R.}~\bibnamefont {Penco}}, \bibinfo {author} {\bibfnamefont {F.}~\bibnamefont {Piazza}},\ and\ \bibinfo {author} {\bibfnamefont {R.}~\bibnamefont {Rattazzi}},\ }\bibfield  {title} {\bibinfo {title} {Zoology of condensed matter: framids, ordinary stuff, extra-ordinary stuff},\ }\href {https://doi.org/10.1007/JHEP06(2015)155} {\bibfield  {journal} {\bibinfo  {journal} {Journal of High Energy Physics}\ }\textbf {\bibinfo {volume} {2015}},\ \bibinfo {pages} {155} (\bibinfo {year} {2015})}\BibitemShut {NoStop}%
\bibitem [{\citenamefont {Geracie}\ \emph {et~al.}(2015)\citenamefont {Geracie}, \citenamefont {Son}, \citenamefont {Wu},\ and\ \citenamefont {Wu}}]{Son2015}%
  \BibitemOpen
  \bibfield  {author} {\bibinfo {author} {\bibfnamefont {M.}~\bibnamefont {Geracie}}, \bibinfo {author} {\bibfnamefont {D.~T.}\ \bibnamefont {Son}}, \bibinfo {author} {\bibfnamefont {C.}~\bibnamefont {Wu}},\ and\ \bibinfo {author} {\bibfnamefont {S.-F.}\ \bibnamefont {Wu}},\ }\bibfield  {title} {\bibinfo {title} {Spacetime symmetries of the quantum {Hall} effect},\ }\href {https://doi.org/10.1103/PhysRevD.91.045030} {\bibfield  {journal} {\bibinfo  {journal} {Physical Review D}\ }\textbf {\bibinfo {volume} {91}},\ \bibinfo {pages} {045030} (\bibinfo {year} {2015})}\BibitemShut {NoStop}%
\bibitem [{\citenamefont {Jensen}(2018)}]{Jensen2018}%
  \BibitemOpen
  \bibfield  {author} {\bibinfo {author} {\bibfnamefont {K.}~\bibnamefont {Jensen}},\ }\bibfield  {title} {\bibinfo {title} {On the coupling of {Galilean}-invariant field theories to curved spacetime},\ }\href {https://doi.org/10.21468/SciPostPhys.5.1.011} {\bibfield  {journal} {\bibinfo  {journal} {SciPost Physics}\ }\textbf {\bibinfo {volume} {5}},\ \bibinfo {pages} {011} (\bibinfo {year} {2018})}\BibitemShut {NoStop}%
\bibitem [{\citenamefont {Son}(2005)}]{Son2005SupersolidEFT}%
  \BibitemOpen
  \bibfield  {author} {\bibinfo {author} {\bibfnamefont {D.~T.}\ \bibnamefont {Son}},\ }\bibfield  {title} {\bibinfo {title} {Effective {Lagrangian} and topological interactions in supersolids},\ }\href {https://doi.org/10.1103/PhysRevLett.94.175301} {\bibfield  {journal} {\bibinfo  {journal} {Physical Review Letters}\ }\textbf {\bibinfo {volume} {94}},\ \bibinfo {pages} {175301} (\bibinfo {year} {2005})}\BibitemShut {NoStop}%
\bibitem [{\citenamefont {Efrati}\ \emph {et~al.}(2009)\citenamefont {Efrati}, \citenamefont {Sharon},\ and\ \citenamefont {Kupferman}}]{non_euc_plates}%
  \BibitemOpen
  \bibfield  {author} {\bibinfo {author} {\bibfnamefont {E.}~\bibnamefont {Efrati}}, \bibinfo {author} {\bibfnamefont {E.}~\bibnamefont {Sharon}},\ and\ \bibinfo {author} {\bibfnamefont {R.}~\bibnamefont {Kupferman}},\ }\bibfield  {title} {\bibinfo {title} {Elastic theory of unconstrained non-{Euclidean} plates},\ }\href {https://doi.org/10.1016/j.jmps.2008.12.004} {\bibfield  {journal} {\bibinfo  {journal} {Journal of the Mechanics and Physics of Solids}\ }\textbf {\bibinfo {volume} {57}},\ \bibinfo {pages} {762} (\bibinfo {year} {2009})}\BibitemShut {NoStop}%
\bibitem [{\citenamefont {Yoo}\ and\ \citenamefont {Dorsey}(2010)}]{yoo_hydrodynamic_2010}%
  \BibitemOpen
  \bibfield  {author} {\bibinfo {author} {\bibfnamefont {C.-D.}\ \bibnamefont {Yoo}}\ and\ \bibinfo {author} {\bibfnamefont {A.~T.}\ \bibnamefont {Dorsey}},\ }\bibfield  {title} {\bibinfo {title} {Hydrodynamic theory of supersolids: {Variational} principle, effective {Lagrangian}, and density-density correlation function},\ }\href {https://doi.org/10.1103/PhysRevB.81.134518} {\bibfield  {journal} {\bibinfo  {journal} {Physical Review B}\ }\textbf {\bibinfo {volume} {81}},\ \bibinfo {pages} {134518} (\bibinfo {year} {2010})}\BibitemShut {NoStop}%
\bibitem [{\citenamefont {Dahlen}(1968)}]{Dahlen1968}%
  \BibitemOpen
  \bibfield  {author} {\bibinfo {author} {\bibfnamefont {F.~A.}\ \bibnamefont {Dahlen}},\ }\bibfield  {title} {\bibinfo {title} {The normal modes of a rotating, elliptical earth},\ }\href {https://doi.org/10.1111/j.1365-246X.1968.tb00229.x} {\bibfield  {journal} {\bibinfo  {journal} {Geophysical Journal International}\ }\textbf {\bibinfo {volume} {16}},\ \bibinfo {pages} {329} (\bibinfo {year} {1968})}\BibitemShut {NoStop}%
\bibitem [{\citenamefont {Sohmen}\ \emph {et~al.}(2021)\citenamefont {Sohmen}, \citenamefont {Politi}, \citenamefont {Klaus}, \citenamefont {Chomaz}, \citenamefont {Mark}, \citenamefont {Norcia},\ and\ \citenamefont {Ferlaino}}]{Sohmen2021Birth}%
  \BibitemOpen
  \bibfield  {author} {\bibinfo {author} {\bibfnamefont {M.}~\bibnamefont {Sohmen}}, \bibinfo {author} {\bibfnamefont {C.}~\bibnamefont {Politi}}, \bibinfo {author} {\bibfnamefont {L.}~\bibnamefont {Klaus}}, \bibinfo {author} {\bibfnamefont {L.}~\bibnamefont {Chomaz}}, \bibinfo {author} {\bibfnamefont {M.~J.}\ \bibnamefont {Mark}}, \bibinfo {author} {\bibfnamefont {M.~A.}\ \bibnamefont {Norcia}},\ and\ \bibinfo {author} {\bibfnamefont {F.}~\bibnamefont {Ferlaino}},\ }\bibfield  {title} {\bibinfo {title} {Birth, life, and death of a dipolar supersolid},\ }\href {https://doi.org/10.1103/PhysRevLett.126.233401} {\bibfield  {journal} {\bibinfo  {journal} {Physical Review Letters}\ }\textbf {\bibinfo {volume} {126}},\ \bibinfo {pages} {233401} (\bibinfo {year} {2021})}\BibitemShut {NoStop}%
\bibitem [{\citenamefont {Cinti}\ \emph {et~al.}(2010)\citenamefont {Cinti}, \citenamefont {Jain}, \citenamefont {Boninsegni}, \citenamefont {Micheli}, \citenamefont {Zoller},\ and\ \citenamefont {Pupillo}}]{Cinti2010Droplet}%
  \BibitemOpen
  \bibfield  {author} {\bibinfo {author} {\bibfnamefont {F.}~\bibnamefont {Cinti}}, \bibinfo {author} {\bibfnamefont {P.}~\bibnamefont {Jain}}, \bibinfo {author} {\bibfnamefont {M.}~\bibnamefont {Boninsegni}}, \bibinfo {author} {\bibfnamefont {A.}~\bibnamefont {Micheli}}, \bibinfo {author} {\bibfnamefont {P.}~\bibnamefont {Zoller}},\ and\ \bibinfo {author} {\bibfnamefont {G.}~\bibnamefont {Pupillo}},\ }\bibfield  {title} {\bibinfo {title} {Supersolid droplet crystal in a dipole-blockaded gas},\ }\href {https://doi.org/10.1103/PhysRevLett.105.135301} {\bibfield  {journal} {\bibinfo  {journal} {Physical Review Letters}\ }\textbf {\bibinfo {volume} {105}},\ \bibinfo {pages} {135301} (\bibinfo {year} {2010})}\BibitemShut {NoStop}%
\bibitem [{\citenamefont {Ciardi}\ \emph {et~al.}(2024)\citenamefont {Ciardi}, \citenamefont {Cinti}, \citenamefont {Pellicane},\ and\ \citenamefont {Prestipino}}]{Ciardi2024BubbleSupersolids}%
  \BibitemOpen
  \bibfield  {author} {\bibinfo {author} {\bibfnamefont {M.}~\bibnamefont {Ciardi}}, \bibinfo {author} {\bibfnamefont {F.}~\bibnamefont {Cinti}}, \bibinfo {author} {\bibfnamefont {G.}~\bibnamefont {Pellicane}},\ and\ \bibinfo {author} {\bibfnamefont {S.}~\bibnamefont {Prestipino}},\ }\bibfield  {title} {\bibinfo {title} {Supersolid phases of bosonic particles in a bubble trap},\ }\href {https://doi.org/10.1103/PhysRevLett.132.026001} {\bibfield  {journal} {\bibinfo  {journal} {Physical Review Letters}\ }\textbf {\bibinfo {volume} {132}},\ \bibinfo {pages} {026001} (\bibinfo {year} {2024})}\BibitemShut {NoStop}%
\bibitem [{\citenamefont {Weckesser}\ \emph {et~al.}(2025)\citenamefont {Weckesser}, \citenamefont {Srakaew}, \citenamefont {Blatz}, \citenamefont {Wei}, \citenamefont {Adler}, \citenamefont {Agrawal}, \citenamefont {Bohrdt}, \citenamefont {Bloch},\ and\ \citenamefont {Zeiher}}]{Weckesser2025Dressed}%
  \BibitemOpen
  \bibfield  {author} {\bibinfo {author} {\bibfnamefont {P.}~\bibnamefont {Weckesser}}, \bibinfo {author} {\bibfnamefont {K.}~\bibnamefont {Srakaew}}, \bibinfo {author} {\bibfnamefont {T.}~\bibnamefont {Blatz}}, \bibinfo {author} {\bibfnamefont {D.}~\bibnamefont {Wei}}, \bibinfo {author} {\bibfnamefont {D.}~\bibnamefont {Adler}}, \bibinfo {author} {\bibfnamefont {S.}~\bibnamefont {Agrawal}}, \bibinfo {author} {\bibfnamefont {A.}~\bibnamefont {Bohrdt}}, \bibinfo {author} {\bibfnamefont {I.}~\bibnamefont {Bloch}},\ and\ \bibinfo {author} {\bibfnamefont {J.}~\bibnamefont {Zeiher}},\ }\bibfield  {title} {\bibinfo {title} {Realization of a {Rydberg}-dressed extended {Bose--Hubbard} model},\ }\href {https://doi.org/10.1126/science.adq7082} {\bibfield  {journal} {\bibinfo  {journal} {Science}\ }\textbf {\bibinfo {volume} {390}},\ \bibinfo {pages} {849} (\bibinfo {year} {2025})}\BibitemShut {NoStop}%
\bibitem [{\citenamefont {Armas}\ \emph {et~al.}(2023)\citenamefont {Armas}, \citenamefont {van Heumen}, \citenamefont {Jain},\ and\ \citenamefont {Lier}}]{ArmasJainLier2023}%
  \BibitemOpen
  \bibfield  {author} {\bibinfo {author} {\bibfnamefont {J.}~\bibnamefont {Armas}}, \bibinfo {author} {\bibfnamefont {E.}~\bibnamefont {van Heumen}}, \bibinfo {author} {\bibfnamefont {A.}~\bibnamefont {Jain}},\ and\ \bibinfo {author} {\bibfnamefont {R.}~\bibnamefont {Lier}},\ }\bibfield  {title} {\bibinfo {title} {Hydrodynamics of plastic deformations in electronic crystals},\ }\href {https://doi.org/10.1103/PhysRevB.107.155108} {\bibfield  {journal} {\bibinfo  {journal} {Physical Review B}\ }\textbf {\bibinfo {volume} {107}},\ \bibinfo {pages} {155108} (\bibinfo {year} {2023})}\BibitemShut {NoStop}%
\bibitem [{\citenamefont {Jensen}(2015)}]{Jensen2015}%
  \BibitemOpen
  \bibfield  {author} {\bibinfo {author} {\bibfnamefont {K.}~\bibnamefont {Jensen}},\ }\bibfield  {title} {\bibinfo {title} {Aspects of hot {Galilean} field theory},\ }\href {https://doi.org/10.1007/JHEP04(2015)123} {\bibfield  {journal} {\bibinfo  {journal} {Journal of High Energy Physics}\ }\textbf {\bibinfo {volume} {2015}},\ \bibinfo {pages} {123} (\bibinfo {year} {2015})}\BibitemShut {NoStop}%
\bibitem [{\citenamefont {Zhou}\ \emph {et~al.}(2026)\citenamefont {Zhou}, \citenamefont {Tsaloukidis}, \citenamefont {Binysh}, \citenamefont {Chen}, \citenamefont {Fakhri}, \citenamefont {Coulais},\ and\ \citenamefont {Sur{\'o}wka}}]{Zhou2025CurvedOdd}%
  \BibitemOpen
  \bibfield  {author} {\bibinfo {author} {\bibfnamefont {Y.}~\bibnamefont {Zhou}}, \bibinfo {author} {\bibfnamefont {L.}~\bibnamefont {Tsaloukidis}}, \bibinfo {author} {\bibfnamefont {J.}~\bibnamefont {Binysh}}, \bibinfo {author} {\bibfnamefont {Y.}~\bibnamefont {Chen}}, \bibinfo {author} {\bibfnamefont {N.}~\bibnamefont {Fakhri}}, \bibinfo {author} {\bibfnamefont {C.}~\bibnamefont {Coulais}},\ and\ \bibinfo {author} {\bibfnamefont {P.}~\bibnamefont {Sur{\'o}wka}},\ }\bibfield  {title} {\bibinfo {title} {Curved odd elasticity},\ }\href {https://doi.org/10.1103/fhwd-lmgk} {\bibfield  {journal} {\bibinfo  {journal} {Physical Review Letters}\ }\textbf {\bibinfo {volume} {137}},\ \bibinfo {pages} {088301} (\bibinfo {year} {2026})}\BibitemShut {NoStop}%
\bibitem [{\citenamefont {Gradshteyn}\ and\ \citenamefont {Ryzhik}(2007)}]{integral_table}%
  \BibitemOpen
  \bibfield  {author} {\bibinfo {author} {\bibfnamefont {I.~S.}\ \bibnamefont {Gradshteyn}}\ and\ \bibinfo {author} {\bibfnamefont {I.~M.}\ \bibnamefont {Ryzhik}},\ }\href@noop {} {\emph {\bibinfo {title} {Table of Integrals, Series, and Products}}},\ \bibinfo {edition} {7th}\ ed.\ (\bibinfo  {publisher} {Academic Press},\ \bibinfo {address} {Amsterdam},\ \bibinfo {year} {2007})\BibitemShut {NoStop}%
\bibitem [{\citenamefont {Poli}\ \emph {et~al.}(2024)\citenamefont {Poli}, \citenamefont {Baillie}, \citenamefont {Ferlaino},\ and\ \citenamefont {Blakie}}]{Poli2024Excitations2dSS}%
  \BibitemOpen
  \bibfield  {author} {\bibinfo {author} {\bibfnamefont {E.}~\bibnamefont {Poli}}, \bibinfo {author} {\bibfnamefont {D.}~\bibnamefont {Baillie}}, \bibinfo {author} {\bibfnamefont {F.}~\bibnamefont {Ferlaino}},\ and\ \bibinfo {author} {\bibfnamefont {P.~B.}\ \bibnamefont {Blakie}},\ }\bibfield  {title} {\bibinfo {title} {Excitations of a two-dimensional supersolid},\ }\href {https://doi.org/10.1103/PhysRevA.110.053301} {\bibfield  {journal} {\bibinfo  {journal} {Physical Review A}\ }\textbf {\bibinfo {volume} {110}},\ \bibinfo {pages} {053301} (\bibinfo {year} {2024})}\BibitemShut {NoStop}%
\bibitem [{\citenamefont {Poli}\ \emph {et~al.}(2026)\citenamefont {Poli}, \citenamefont {Martone}, \citenamefont {Stringari},\ and\ \citenamefont {Recati}}]{2026arXiv260401751P}%
  \BibitemOpen
  \bibfield  {author} {\bibinfo {author} {\bibfnamefont {E.}~\bibnamefont {Poli}}, \bibinfo {author} {\bibfnamefont {G.~I.}\ \bibnamefont {Martone}}, \bibinfo {author} {\bibfnamefont {S.}~\bibnamefont {Stringari}},\ and\ \bibinfo {author} {\bibfnamefont {A.}~\bibnamefont {Recati}},\ }\href {https://doi.org/10.48550/arXiv.2604.01751} {\bibinfo {title} {Sound propagation in striped supersolid cold gases at zero temperature}} (\bibinfo {year} {2026}),\ \Eprint {https://arxiv.org/abs/2604.01751} {arXiv:2604.01751 [cond-mat.quant-gas]} \BibitemShut {NoStop}%
\bibitem [{\citenamefont {Jia}\ \emph {et~al.}(2022)\citenamefont {Jia}, \citenamefont {Huang}, \citenamefont {Qiu}, \citenamefont {Zhou}, \citenamefont {Yan},\ and\ \citenamefont {Wang}}]{Jia2022ShellExpansion}%
  \BibitemOpen
  \bibfield  {author} {\bibinfo {author} {\bibfnamefont {F.}~\bibnamefont {Jia}}, \bibinfo {author} {\bibfnamefont {Z.}~\bibnamefont {Huang}}, \bibinfo {author} {\bibfnamefont {L.}~\bibnamefont {Qiu}}, \bibinfo {author} {\bibfnamefont {R.}~\bibnamefont {Zhou}}, \bibinfo {author} {\bibfnamefont {Y.}~\bibnamefont {Yan}},\ and\ \bibinfo {author} {\bibfnamefont {D.}~\bibnamefont {Wang}},\ }\bibfield  {title} {\bibinfo {title} {Expansion dynamics of a shell-shaped {Bose--Einstein} condensate},\ }\href {https://doi.org/10.1103/PhysRevLett.129.243402} {\bibfield  {journal} {\bibinfo  {journal} {Physical Review Letters}\ }\textbf {\bibinfo {volume} {129}},\ \bibinfo {pages} {243402} (\bibinfo {year} {2022})}\BibitemShut {NoStop}%
\bibitem [{\citenamefont {Veyron}\ \emph {et~al.}(2025)\citenamefont {Veyron}, \citenamefont {M{\'e}tayer}, \citenamefont {G{\'e}rent}, \citenamefont {Huang}, \citenamefont {Beraud}, \citenamefont {Garraway}, \citenamefont {Bernon},\ and\ \citenamefont {Battelier}}]{Veyron2025AllOptical}%
  \BibitemOpen
  \bibfield  {author} {\bibinfo {author} {\bibfnamefont {R.}~\bibnamefont {Veyron}}, \bibinfo {author} {\bibfnamefont {C.}~\bibnamefont {M{\'e}tayer}}, \bibinfo {author} {\bibfnamefont {J.-B.}\ \bibnamefont {G{\'e}rent}}, \bibinfo {author} {\bibfnamefont {R.}~\bibnamefont {Huang}}, \bibinfo {author} {\bibfnamefont {E.}~\bibnamefont {Beraud}}, \bibinfo {author} {\bibfnamefont {B.~M.}\ \bibnamefont {Garraway}}, \bibinfo {author} {\bibfnamefont {S.}~\bibnamefont {Bernon}},\ and\ \bibinfo {author} {\bibfnamefont {B.}~\bibnamefont {Battelier}},\ }\href {https://doi.org/10.1116/5.0305448} {\bibinfo {title} {All-optical bubble trap for ultracold atoms in microgravity}} (\bibinfo {year} {2025}),\ \Eprint {https://arxiv.org/abs/2510.05734} {arXiv:2510.05734 [cond-mat.quant-gas]} \BibitemShut {NoStop}%
\bibitem [{\citenamefont {Hertkorn}\ \emph {et~al.}(2021)\citenamefont {Hertkorn}, \citenamefont {Schmidt}, \citenamefont {Guo}, \citenamefont {B{\"o}ttcher}, \citenamefont {Ng}, \citenamefont {Graham}, \citenamefont {Uerlings}, \citenamefont {Langen}, \citenamefont {Zwierlein},\ and\ \citenamefont {Pfau}}]{Hertkorn2021Pattern}%
  \BibitemOpen
  \bibfield  {author} {\bibinfo {author} {\bibfnamefont {J.}~\bibnamefont {Hertkorn}}, \bibinfo {author} {\bibfnamefont {J.-N.}\ \bibnamefont {Schmidt}}, \bibinfo {author} {\bibfnamefont {M.}~\bibnamefont {Guo}}, \bibinfo {author} {\bibfnamefont {F.}~\bibnamefont {B{\"o}ttcher}}, \bibinfo {author} {\bibfnamefont {K.~S.~H.}\ \bibnamefont {Ng}}, \bibinfo {author} {\bibfnamefont {S.~D.}\ \bibnamefont {Graham}}, \bibinfo {author} {\bibfnamefont {P.}~\bibnamefont {Uerlings}}, \bibinfo {author} {\bibfnamefont {T.}~\bibnamefont {Langen}}, \bibinfo {author} {\bibfnamefont {M.}~\bibnamefont {Zwierlein}},\ and\ \bibinfo {author} {\bibfnamefont {T.}~\bibnamefont {Pfau}},\ }\bibfield  {title} {\bibinfo {title} {Pattern formation in quantum ferrofluids: From supersolids to superglasses},\ }\href {https://doi.org/10.1103/PhysRevResearch.3.033125} {\bibfield  {journal} {\bibinfo  {journal} {Physical Review Research}\ }\textbf {\bibinfo {volume} {3}},\ \bibinfo {pages} {033125} (\bibinfo {year} {2021})}\BibitemShut {NoStop}%
\bibitem [{\citenamefont {Biagioni}\ \emph {et~al.}(2024)\citenamefont {Biagioni}, \citenamefont {Antolini}, \citenamefont {Donelli}, \citenamefont {Pezz{\`e}}, \citenamefont {Smerzi}, \citenamefont {Fattori}, \citenamefont {Fioretti}, \citenamefont {Gabbanini}, \citenamefont {Inguscio}, \citenamefont {Tanzi},\ and\ \citenamefont {Modugno}}]{Biagioni2024Josephson}%
  \BibitemOpen
  \bibfield  {author} {\bibinfo {author} {\bibfnamefont {G.}~\bibnamefont {Biagioni}}, \bibinfo {author} {\bibfnamefont {N.}~\bibnamefont {Antolini}}, \bibinfo {author} {\bibfnamefont {B.}~\bibnamefont {Donelli}}, \bibinfo {author} {\bibfnamefont {L.}~\bibnamefont {Pezz{\`e}}}, \bibinfo {author} {\bibfnamefont {A.}~\bibnamefont {Smerzi}}, \bibinfo {author} {\bibfnamefont {M.}~\bibnamefont {Fattori}}, \bibinfo {author} {\bibfnamefont {A.}~\bibnamefont {Fioretti}}, \bibinfo {author} {\bibfnamefont {C.}~\bibnamefont {Gabbanini}}, \bibinfo {author} {\bibfnamefont {M.}~\bibnamefont {Inguscio}}, \bibinfo {author} {\bibfnamefont {L.}~\bibnamefont {Tanzi}},\ and\ \bibinfo {author} {\bibfnamefont {G.}~\bibnamefont {Modugno}},\ }\bibfield  {title} {\bibinfo {title} {Measurement of the superfluid fraction of a supersolid by {Josephson} effect},\ }\href {https://doi.org/10.1038/s41586-024-07361-9} {\bibfield  {journal} {\bibinfo  {journal} {Nature}\ }\textbf {\bibinfo {volume} {629}},\ \bibinfo {pages} {773} (\bibinfo
  {year} {2024})}\BibitemShut {NoStop}%
\bibitem [{\citenamefont {Senarath~Yapa}\ and\ \citenamefont {Bland}(2025)}]{SenarathYapa2025Shear}%
  \BibitemOpen
  \bibfield  {author} {\bibinfo {author} {\bibfnamefont {P.}~\bibnamefont {Senarath~Yapa}}\ and\ \bibinfo {author} {\bibfnamefont {T.}~\bibnamefont {Bland}},\ }\href {https://doi.org/10.1103/xz57-52ft} {\bibinfo {title} {Anomalous dispersion of shear waves in dipolar supersolids}} (\bibinfo {year} {2025}),\ \Eprint {https://arxiv.org/abs/2410.16060} {arXiv:2410.16060 [cond-mat.quant-gas]} \BibitemShut {NoStop}%
\bibitem [{\citenamefont {Ripley}\ \emph {et~al.}(2023)\citenamefont {Ripley}, \citenamefont {Baillie},\ and\ \citenamefont {Blakie}}]{Ripley2023Planar}%
  \BibitemOpen
  \bibfield  {author} {\bibinfo {author} {\bibfnamefont {B.~T.~E.}\ \bibnamefont {Ripley}}, \bibinfo {author} {\bibfnamefont {D.}~\bibnamefont {Baillie}},\ and\ \bibinfo {author} {\bibfnamefont {P.~B.}\ \bibnamefont {Blakie}},\ }\bibfield  {title} {\bibinfo {title} {Two-dimensional supersolidity in a planar dipolar {Bose} gas},\ }\href {https://doi.org/10.1103/PhysRevA.108.053321} {\bibfield  {journal} {\bibinfo  {journal} {Physical Review A}\ }\textbf {\bibinfo {volume} {108}},\ \bibinfo {pages} {053321} (\bibinfo {year} {2023})}\BibitemShut {NoStop}%
\bibitem [{\citenamefont {Macr{\`i}}\ \emph {et~al.}(2013)\citenamefont {Macr{\`i}}, \citenamefont {Maucher}, \citenamefont {Cinti},\ and\ \citenamefont {Pohl}}]{Macri2013SoftCore}%
  \BibitemOpen
  \bibfield  {author} {\bibinfo {author} {\bibfnamefont {T.}~\bibnamefont {Macr{\`i}}}, \bibinfo {author} {\bibfnamefont {F.}~\bibnamefont {Maucher}}, \bibinfo {author} {\bibfnamefont {F.}~\bibnamefont {Cinti}},\ and\ \bibinfo {author} {\bibfnamefont {T.}~\bibnamefont {Pohl}},\ }\bibfield  {title} {\bibinfo {title} {Elementary excitations of ultracold soft-core bosons across the superfluid--supersolid phase transition},\ }\href {https://doi.org/10.1103/PhysRevA.87.061602} {\bibfield  {journal} {\bibinfo  {journal} {Physical Review A}\ }\textbf {\bibinfo {volume} {87}},\ \bibinfo {pages} {061602} (\bibinfo {year} {2013})}\BibitemShut {NoStop}%
\end{thebibliography}%
 
\newpage
\phantom{a}
\newpage
\setcounter{figure}{0}
\renewcommand{\thefigure}{S\arabic{figure}}
\renewcommand{\theHfigure}{S\arabic{figure}}
\setcounter{equation}{0}
\renewcommand{\theequation}{S\arabic{equation}}
\renewcommand{\theHequation}{S\arabic{equation}}
\setcounter{section}{0}

\onecolumngrid
\begin{center}
{\large \bf Supplementary Material:\\
Curvature Gradients Turn Shear into Superflow in Supersolids}\\
\vspace{0.3cm}
\end{center}
\setcounter{page}{1}
 
%
%
%
%
\section{Basics of Newton-Cartan and lattice description}
\label{sec: SM NC crystal}
 
This section collects the geometric background: the Newton-Cartan (NC) structure on which non-relativistic theories are defined covariantly, the description of a crystal by Goldstone fields, and how a supersolid is coupled to both.
 
\subsection{Newton-Cartan geometry}
\label{sec: SM NC}
 
Non-relativistic field theories can be coupled to curved backgrounds and external gauge fields, keeping every symmetry manifest, by placing them on a Newton-Cartan structure \cite{Son2015,Jensen2018}. The spacetime is a $(d+1)$-dimensional manifold $\mcM$ with a clock form $n_\mu$ and a symmetric tensor $h^{\mu\nu}$ of rank $d$, with $n_\mu h^{\mu\nu}=0$. Since $h^{\mu\nu}$ is degenerate, we also need a vector field $v^\mu$ with $n_\mu v^\mu=1$, which defines implicitly the spatial metric $h_{\mu\nu}$ and the projector $P^\nu_\mu$ onto directions transverse to $v^\mu$,
\begin{equation}
    h_{\mu\nu}v^\nu=0, \qquad h_{\mu\rho}h^{\nu\rho}=P^\nu_\mu \equiv \dl^\nu_\mu - v^\nu n_\mu.
    \label{eq: SM NC data}
\end{equation}
We use the connection compatible with $(n_\mu,h^{\mu\nu})$ that has purely temporal torsion \cite{Jensen2018},
\begin{equation}
    \Gm^\mu_{\nu\rho} = v^\mu \partial_\rho n_\nu + \frac{1}{2}h^{\mu\sigma}\lt(\partial_\nu h_{\sigma\rho}+\partial_\rho h_{\nu\sigma}-\partial_\sigma h_{\nu\rho}\rt) + h^{\mu\sigma}n_{(\nu}F_{\rho)\sigma}, \qquad
    T^\mu_{\nu\rho} = v^\mu F^n_{\rho\nu}, \qquad F^n_{\mu\nu}\coloneqq2\partial_{[\mu}n_{\nu]},
    \label{eq: SM NC connection}
\end{equation}
with covariant derivative $\bmD_\mu$. Consistency of this connection requires $dF=0$, so locally $F=dA$, with $A_\mu$ a $U(1)$ gauge field. The background data is then $(\mcM,n_\mu,h^{\mu\nu},v^\mu,A_\mu)$, and volumes are measured with $\sqrt{\gm}$, where $\gm_{\mu\nu}\coloneqq n_\mu n_\nu + h_{\mu\nu}$.
 
The vector field $v^\mu$ is not unique. For any one-form $\psi_\mu$, $v'^\mu=v^\mu+h^{\mu\nu}\psi_\nu$ also satisfies $v'^\mu n_\mu=1$, and this redundancy is a \emph{Milne boost}. Keeping the connection Milne invariant fixes how $h_{\mu\nu}$ and $A_\mu$ shift,
\begin{equation}
    h'_{\mu\nu} = h_{\mu\nu} - \lt(n_\mu P^\rho_\nu + n_\nu P^\rho_\mu\rt)\psi_\rho + n_\mu n_\nu h^{\rho\sigma}\psi_\rho\psi_\sigma, \qquad A'_\mu = A_\mu + P^\nu_\mu\psi_\nu - \frac{1}{2}n_\mu h^{\nu\rho}\psi_\nu\psi_\rho.
    \label{eq: SM Milne shifts}
\end{equation}
Jensen's proposal \cite{Jensen2018} is that a Galilean-invariant field theory coupled to NC geometry has a generating functional invariant under diffeomorphisms, $U(1)$ gauge transformations and Milne boosts. Around flat space this reproduces the Galilean algebra, with the $U(1)$ charge coupled to $A_\mu$ as the central element. Therefore, $A_\mu$ geometrizes particle-number conservation.
 
\subsection{Crystal geometry}
\label{sec: SM crystal}
 
A crystal is described by $d$ one-forms $e^A_\mu$, one for each lattice direction. For them to define lattice planes, the distribution they annihilate must be integrable, and by the Frobenius theorem the frame can then be chosen exact \cite{ArmasJainLier2023},
\begin{equation}
    e^A_\mu=\partial_\mu X^A .
\end{equation}
The scalars $X^A$ label the lattice sites and act as Lagrangian coordinates. In flat space they are the Goldstone fields of the broken translations; on a curved surface, where there are no translations to break, they remain the fields whose fluctuations are the phonons. More precisely, $X\colon\Sigma\to\mathcal B$ maps the spatial slice to a comoving manifold $\mathcal B$, and $X^A$ are its components in a chart of $\mathcal B$. By reparametrization invariance nothing depends on the chart, so the fact that a closed surface needs more than one chart is immaterial. On a closed surface the crystal cannot be free of defects: a triangular lattice on a sphere carries a total disclination charge of $12$. We coarse-grain over the defects and treat them as frozen on the time scales of the collective modes, so that they neither move nor are created. Their density then enters only through the reference metric $G_{AB}$ of $\mathcal B$, whose Gaussian curvature is the coarse-grained disclination density. The choice~\eqref{eq: SM stress free} assumes that this density screens the curvature of the substrate, so that the coarse-grained lattice is stress free. The description holds at wavelengths long compared with the lattice spacing, and for crystals with enough sites that the defect density can be treated as smooth. The metric $h^{\mu\nu}$ induces a metric on the comoving space,
\begin{equation}
    U^{AB} = h^{\mu\nu}e^A_\mu e^B_\nu, \qquad U_{AB}=(U^{AB})^{-1},
    \label{eq: SM induced metric}
\end{equation}
and elastic deformations are measured with respect to a \emph{reference} metric $G_{AB}$, the metric the undeformed lattice would carry,
\begin{equation}
    k_{AB}=\frac{1}{2}(G_{AB}-U_{AB}),
    \label{eq: SM strain}
\end{equation}
with $k_{AB}$ the strain. The labels of the lattice sites carry no physical meaning, so the theory is invariant under reparametrizations $X^A\to X'^A(X)$, under which $U^{AB}$ and $G_{AB}$ transform as tensors on comoving space. This is why $G_{AB}$ is needed. With $U^{AB}$ alone no scalar can be formed, since even $\det U^{AB}$ picks up the square of the Jacobian. The invariants are those of $U^{AB}G_{BC}$; for instance, $\sqrt{\det(U^{AB}G_{BC})}$ is the density of lattice sites relative to the reference. In flat space $G_{AB}=\delta_{AB}$ is usually left implicit in the contractions, and only the relabelings that preserve it are allowed.
 
\subsection{Coupling a supersolid to Newton Cartan}
\label{sec: SM coupling}
 
A supersolid breaks the $U(1)$ particle-number symmetry and spatial translations at once. Its Goldstone fields are the phase $\theta$ and the comoving coordinates $X^A$. In flat space, Galilean invariance requires the momentum density to equal the mass current, $T^{0i}=\mass j^i$, and Son builds this in by letting $\theta$ and $X^A$ enter only through \cite{Son2005SupersolidEFT,yoo_hydrodynamic_2010}
\begin{equation}
    \mu = \partial_t\theta - \frac{(\partial_i\theta)^2}{2\mass}, \qquad w^A = -\partial_t X^A + \frac{1}{\mass}\partial_i\theta\,\partial_iX^A, \qquad u^{AB}=\partial_iX^A\partial_iX^B.
    \label{eq: SM Son variables}
\end{equation}
Our variables are the covariant versions of these,
\begin{equation}
    \mu = v^\mu\xi_\mu - \frac{1}{2\mass}h^{\mu\nu}\xi_\mu\xi_\nu, \qquad W^A = -v^\mu e^A_\mu + \frac{1}{\mass}h^{\mu\nu}\xi_\mu e^A_\nu, \qquad U^{AB} = h^{\mu\nu}e^A_\mu e^B_\nu, \qquad \xi_\mu = \partial_\mu\theta + \mass A_\mu,
    \label{eq: SM variables}
\end{equation}
where $\xi_\mu$ is gauge invariant, since $\theta\to\theta-\mass\Lambda$ when $A_\mu\to A_\mu+\partial_\mu\Lambda$. Here $\mu$ is the local chemical potential, $W^A$ the counterflow and $U^{AB}$ the induced metric. On the flat background $n_\mu dx^\mu=dt$, $h^{\mu\nu}=\delta^{ij}$, $v^\mu=\partial_t$, $A=0$, they reduce to \eqref{eq: SM Son variables}. The advantage is that Galilean invariance becomes Milne invariance. Under a boost,
\begin{equation}
    v^\mu \to v^\mu + \psi^\mu, \qquad \xi_\mu \to \xi_\mu + \mass\big(P^\nu_\mu\psi_\nu - \tfrac12 n_\mu\psi^2\big),
    \label{eq: SM boost of v and xi}
\end{equation}
with $\psi^\mu\coloneqq h^{\mu\nu}\psi_\nu$ and $\psi^2\coloneqq h^{\mu\nu}\psi_\mu\psi_\nu$, while $h^{\mu\nu}$ and $e^A_\mu$ do not change. Then $\mu$, $W^A$, $U^{AB}$ and $\sqrt{\gm}$ are invariant, and any function $F(\mu,W^A,U^{AB};G^{AB}(X))$ is Galilean invariant without further constraints. In Sec.~\ref{sec: SM GF formalism} we will see that $T^{0i}=\mass j^i$ becomes the Milne Ward identity $\mcP_\mu=h_{\mu\nu}\mcJ^\nu$, which holds identically.
 
The velocity of the lattice $u^\mu_X$ is fixed by
\begin{equation}
    u^\mu_X e^A_\mu = 0, \qquad u^\mu_X n_\mu = 1,
    \label{eq: SM Lin constraint}
\end{equation}
which is the covariant form of Lin's constraint: in flat space it reads $(\partial_t+u^i_X\partial_i)X^A=0$, so the comoving coordinates are dragged along by the lattice. The inverse frame obeys $e^\mu_A e^A_\nu = \dl^\mu_\nu - u^\mu_X n_\nu$, with $e^\mu_A = h^{\mu\nu}U_{AB}e^B_\nu$. With the superfluid velocity $u_s^\mu\coloneqq v^\mu-h^{\mu\nu}\xi_\nu/\mass$, which is also Milne invariant, the counterflow reads
\begin{equation}
    W^A = -u_s^\mu e^A_\mu = (u_X^\mu - u_s^\mu)\,e^A_\mu,
    \label{eq: SM W as relative velocity}
\end{equation}
the velocity of the lattice relative to the superfluid, pulled back to comoving space.
\section{Generating Functional Formalism}
\label{sec: SM GF formalism}
 
The Ward identities of the supersolid follow from the generating-functional method used for NC superfluids \cite{Jensen2018,Jensen2015}, now with the lattice Goldstones $X^A$. We record only what differs from the superfluid case.
 
\subsection{Variation and currents}
 
Let $W[n_\mu,h^{\mu\nu},v^\mu,A_\mu,\theta,X^A]$ be the generating functional. Since $(n_\mu,h^{\mu\nu},v^\mu)$ obey \eqref{eq: SM NC data}, we vary $n_\mu$ freely and use the transverse variations $\delta\ovv^\mu$ and $\delta\ovh^{\mu\nu}$, defined by
\begin{equation}
    \dl v^\mu = -v^\mu v^\nu \dl n_\nu + P^\mu_\nu \dl \ovv^\nu, \qquad
    \dl h^{\mu \nu} = -(v^\mu h^{\nu \rho} + v^\nu h^{\mu \rho})\dl n_\rho + P^{\mu}_\rho P^\nu_\sigma \dl \ovh^{\rho \sigma}.
\end{equation}
The currents are then defined by
\begin{equation}
    \delta W = \int d^dx\sqrt{\gm}\Big[\mcJ^\mu \delta A_\mu + U\delta\theta + K_A\delta X^A - \mcP_\mu \delta \ovv^\mu - \mcE^\mu\delta n_\mu - \frac{\delta \ovh^{\mu\nu}}{2}\bmT_{\mu\nu}\Big],
    \label{eq: SM dW supersolid}
\end{equation}
with $\mcJ^\mu$ the particle current, $U$ the source conjugate to the phase, $\mcP_\mu$ the momentum density, $\mcE^\mu$ the energy current and $\bmT_{\mu\nu}$ the stress. The new term is $K_A\delta X^A$, with $K_A$ the force conjugate to the lattice displacement. Covariance under reparametrizations of $X^A$ fixes it to enter linearly in $\delta X^A$ \cite{ArmasJainLier2023}.
 
\subsection{Symmetries and the boundary-term lemma}
 
Diffeomorphisms $\chi^\mu$, Milne boosts $\psi_\mu$ and $U(1)$ gauge transformations $\Lm$ act as
\begin{equation}
    \begin{split}
        \delta n_\mu &= \mcL_\chi n_\mu, \qquad \delta \ovv^\mu = \mcL_\chi v^\mu + h^{\mu\nu}\psi_\nu, \qquad \delta \ovh^{\mu\nu} = \mcL_\chi h^{\mu\nu},\\
        \delta A_\mu &= \mcL_\chi A_\mu + P^\nu_\mu\psi_\nu + \partial_\mu\Lm, \qquad \delta\theta = \mcL_\chi\theta - \mass\Lm, \qquad \delta X^A = \mcL_\chi X^A.
    \end{split}
    \label{eq: SM symmetry variations supersolid}
\end{equation}
The Ward identities follow from the invariance of $W$ for arbitrary $\chi,\psi,\Lm$. They are local, since for any vector $\mfv^\mu$
\begin{equation}
    \Xi_\mu \mfv^\mu \coloneqq (\bmD_\mu - \bmG_\mu)\mfv^\mu = \frac{1}{\sqrt{\gm}}\partial_\mu(\sqrt{\gm}\,\mfv^\mu), \qquad \bmG_\mu\coloneqq T^\nu_{\mu\nu} = -F^n_{\mu\nu}v^\nu,
    \label{eq: SM boundary lemma}
\end{equation}
so $\int d^dx\sqrt{\gm}\,\Xi_\mu\mfv^\mu$ is a boundary term.
 
\subsection{Ward identities}
 
$U(1)$ invariance and Milne boosts give
\begin{equation}
    \Xi_\mu \mcJ^\mu = -\mass U, \qquad \mcP_\mu = h_{\mu\nu}\mcJ^\nu,
    \label{eq: SM U1 Milne WI}
\end{equation}
and diffeomorphisms, split along $v^\mu$ and transverse to it,
\begin{equation}
    \begin{split}
        \Xi_\mu\mcE^\mu &= v^\mu\big(F^n_{\mu\nu}\mcE^\nu - F_{\mu\nu}\mcJ^\nu - U\xi_\mu - K_A e^A_\mu\big) - \bmT_{\mu\nu}\bmD^{(\mu}v^{\nu)},\\
        \Xi_\mu \bmT^{\mu\nu} &= \mcJ^\mu F^\nu{}_\mu + U\xi^\nu + K_A e^{A\nu} - \mcE^\mu (F^n)^\nu{}_\mu + v^\mu\bmD^\nu\mcP_\mu - \bmD_\mu(v^\mu\mcP^\nu),
    \end{split}
    \label{eq: SM E T WI supersolid}
\end{equation}
where indices on $\xi_\mu$ and $e^A_\mu$ are raised with $h^{\mu\nu}$. Compared to the superfluid \cite{Jensen2015}, which has $K_A=0$, the lattice force enters the energy and stress balance in the same slot as the phase source $U$.
 
\subsection{Currents under Milne boosts}
 
Under a boost the variations shift as
\begin{equation}
    \dl A'_\mu=\dl A_\mu - n_\mu\lt(\dl\ovv^\nu\psi_\nu-\frac{\dl\ovh^{\nu\rho}}{2}\psi_\nu\psi_\rho\rt),\qquad
    \dl \ovv'^\mu = \dl \ovv^\mu +\dl \ovh^{\mu\nu}\psi_\nu,
\end{equation}
and therefore
\begin{equation}
    \mcJ'^\mu = \mcJ^\mu, \qquad U'=U, \qquad K_A'=K_A,\qquad
    \mcP'_\mu = \mcP_\mu - \psi_\mu n_\nu\mcJ^\nu,\qquad
    \bmT'^{\mu\nu} = \bmT^{\mu\nu} - 2\mcP^{(\mu}h^{\nu)\sigma}\psi_\sigma + h^{\mu\rho}h^{\nu\kappa}\psi_\rho\psi_\kappa\,n_\sigma\mcJ^\sigma.
    \label{eq: SM current Milne shifts supersolid}
\end{equation}
The particle current, the phase source and the lattice force are Milne invariant on their own.
 
\subsection{Milne-invariant theory}
 
The improved stress tensor
\begin{equation}
    \mcT^{\mu\nu} \coloneqq \bmT^{\mu\nu} + 2\mcP^{(\mu}v^{\nu)} + v^\mu v^\nu n_\sigma\mcJ^\sigma, \qquad n_\nu\mcT^{\mu\nu} = \mcJ^\mu,
    \label{eq: SM improved stress supersolid}
\end{equation}
is Milne invariant, and with it \eqref{eq: SM E T WI supersolid} becomes
\begin{equation}
    \begin{split}
        \Xi_\nu\mcT^{\mu\nu} &= U(\xi^\mu - \mass v^\mu) - \mcE^\nu(F^n)^\mu{}_\nu + K_A e^{A\mu},\\
        \Xi_\mu\mcE^\mu &= \bmG_\mu\mcE^\mu - U\xi_\nu v^\nu - K_A v^\nu e^A_\nu - h_{\rho(\mu}\bmD_{\nu)}v^\rho\,\mcT^{\mu\nu},
    \end{split}
    \label{eq: SM Milne-inv WI supersolid}
\end{equation}
from which the energy current shifts as $\mcE'^\mu = \mcE^\mu - \big(P^\rho_\nu\psi_\rho - \tfrac12 n_\nu\psi^2\big)\mcT^{\mu\nu}$.
 
For the supersolid we can go one step further. The lattice velocity $u^\mu_X$ is already Milne invariant, since it is fixed by $e^A_\mu$ and $n_\mu$ alone. With $u_{X\mu}\coloneqq h_{\mu\nu}u_X^\nu$ and $u_X^2\coloneqq h_{\mu\nu}u_X^\mu u_X^\nu$, the combination $-u_{X\mu}+\frac12 n_\mu u_X^2$ shifts exactly as $A_\mu$, so
\begin{equation}
    \tdh_{\mu\nu} = h_{\mu\nu} - 2u_{X(\mu}n_{\nu)} + u_X^2\, n_\mu n_\nu, \qquad \tdA_\mu = A_\mu + u_{X\mu} - \frac{1}{2}n_\mu u_X^2,
    \label{eq: SM tilde frame supersolid}
\end{equation}
are Milne invariant, together with their connection, of torsion $\tilde T^\mu_{\nu\rho} = u_X^\mu F^n_{\rho\nu}$. In this frame the identities become
\begin{equation}
    (\tilde\Xi_\mu - \tilde{\bmG}_\mu)\tilde{\mcE}^\mu = -U\,\tilde\xi_\nu u_X^\nu - \mcT^{\mu\nu}\tdh_{\rho(\mu}\tilde{\bmD}_{\nu)}u_X^\rho, \qquad
    \tilde\Xi_\nu\mcT^{\mu\nu} = -(F^n)^\mu{}_\nu\tilde{\mcE}^\nu + U(\tilde\xi^\mu - \mass u_X^\mu) + K_A e^{\mu A},
    \label{eq: SM final energy WI}
\end{equation}
with $\tilde{\mcE}^\mu = \mcE^\mu - \big(u_{X\nu} - \frac{1}{2}n_\nu u_X^2\big)\mcT^{\mu\nu}$. Since $u^\mu_X e^A_\mu=0$, the lattice force drops out of the energy identity: it does no work in the frame that moves with the lattice.

\section{Constitutive relations}
\label{sec: SM constitutive}
\label{sec: SM CR}
 
At leading order in derivatives the action is
\begin{equation}
    S=\int d^dx\sqrt{\gm}\,F\big(\mu,W^A,U^{AB};G_{AB}(X)\big),
    \label{eq: SM action supersolid}
\end{equation}
where $F$ is the pressure. Besides the dependence through $e^A_\mu$ in $W^A$ and $U^{AB}$, $F$ depends on $X^A$ explicitly, through the reference metric $G_{AB}(X)$. We define the mass density, the counterflow momentum and the elastic stress,
\begin{equation}
    \rho\coloneqq \mass\frac{\partial F}{\partial\mu}, \qquad p_\mu\coloneqq\frac{\partial F}{\partial W^A}e^A_\mu, \qquad \sigma_{AB}\coloneqq-2\frac{\partial F}{\partial U^{AB}}.
\end{equation}
Varying \eqref{eq: SM action supersolid} with respect to the background at fixed $\theta$ and $X^A$, and comparing with \eqref{eq: SM dW supersolid}, we get
\begin{equation}
    \begin{split}
        \mcJ^\mu &= \rho\,u_s^\mu + p^\mu,\qquad
        \mcP_\mu = P^\nu_\mu\Big(p_\nu - \frac{\rho}{\mass}\xi_\nu\Big),\\
        \mcE^\mu &= -F v^\mu + \frac{v^\nu\xi_\nu}{\mass}\,\mcJ^\mu - (v^\nu p_\nu)\,u_s^\mu - \sigma_{AB}\,(v^\nu e^A_\nu)\,e^{B\mu},\\
        \bmT_{\mu\nu} &= F h_{\mu\nu} + P^\rho_\mu P^\sigma_\nu\Big(\frac{\rho}{\mass^2}\xi_\rho\xi_\sigma - \frac{2}{\mass}\xi_{(\rho}p_{\sigma)} + \sigma_{AB}\,e^A_\rho e^B_\sigma\Big),
    \end{split}
    \label{eq: SM constitutive relations}
\end{equation}
with $p^\mu\coloneqq h^{\mu\nu}p_\nu$. The improved stress tensor is then manifestly Milne invariant,
\begin{equation}
    \mcT^{\mu\nu} = \rho\,u_s^\mu u_s^\nu + 2p^{(\mu}u_s^{\nu)} + F h^{\mu\nu} + \sigma_{AB}\,e^{A\mu}e^{B\nu}.
    \label{eq: SM improved stress constitutive}
\end{equation}

These relations pass three checks, which are the first benchmark of the construction. First, $h_{\mu\nu}\mcJ^\nu=\mcP_\mu$ holds identically, so Son's $T^{0i}=\mass j^i$ follows from the Milne Ward identity. Second, time reversal sends $W^A\to-W^A$ and leaves $\mu$ and $U^{AB}$ invariant, so $F$ is even in $W^A$ and $p_\mu=0$ at rest. Then $\mcJ^0=\rho$, $\mcE^0=\mu\rho/\mass-F$ and $\bmT_{ij}=F\delta_{ij}+\sigma_{AB}e^A_ie^B_j$, which is the thermodynamics of a solid with pressure $F$. Third, for $F\simeq F_0(\mu,U)+\frac12\rho_n(\mu,U)\,U_{AB}W^AW^B$, Eq.~\eqref{eq: SM W as relative velocity} gives $p^\mu=\rho_n(u_X^\mu-u_s^\mu)$, and
\begin{equation}
    \mcJ^\mu = (\rho-\rho_n)\,u_s^\mu + \rho_n\,u_X^\mu ,
\end{equation}
the two-fluid form, with $\rho_s=\rho-\rho_n$ following from the action.

\section{Quadratic action and linear dynamics}
\label{sec: SM quadratic}
 
For collective modes we need the action \eqref{eq: SM action supersolid} only to quadratic order around equilibrium. In this section we fix it by symmetry and derive the equations of motion.
 
\subsection{Equilibrium and allowed terms}
\label{sec: SM equilibrium}
\label{sec: SM allowed terms}
 
We take a static substrate, generated by $v^\mu$, and an equilibrium where both components are at rest in the frame of $v^\mu$,
\begin{equation}
    v^\mu e^A_{0\mu}=0, \qquad \xi_{0\mu}=\mu_0\,n_\mu ,
    \label{eq: SM equilibrium}
\end{equation}
so $u_X^\mu=u_s^\mu=v^\mu$ and $W_0^A=0$. We also take the equilibrium to be stress free. A self-assembled supersolid relaxes to its own unstressed configuration, so we choose the reference metric to be the equilibrium induced metric,
\begin{equation}
    G_{AB}=U_{0AB}, \qquad k_{0AB}=0 .
    \label{eq: SM stress free}
\end{equation}
This is possible on any substrate, whatever its curvature. Since an isometry preserves the curvature, it requires $G_{AB}$ to carry the curvature of the substrate: a flat reference metric on a sphere would leave the equilibrium strained. A curved $G_{AB}$ has non-constant components, so $F$ depends on $X^A$ explicitly, and the internal translations $X^A\to X^A+c^A$ are broken down to the isometries of $G_{AB}$, as befits a non-Euclidean plate \cite{non_euc_plates}. By definition $G_{AB}$ is the metric of the unstressed material at $\mu_0$, so $F$ has no term linear in $k_{AB}$, and \eqref{eq: SM stress free} makes the equilibrium unstrained.
 
Near equilibrium, $F$ depends on $\delta\mu=\mu-\mu_0$, $W^A$ and $k_{AB}$. Time reversal makes $W^A$ appear in pairs, which removes the couplings $W^A\delta\mu$ and $W^Ak_{BC}$. This relies on $W_0^A=0$, and fails for a rotating supersolid or one with a persistent current. For an isotropic crystal the only available tensor is $G_{AB}$, which in two dimensions covers the triangular lattice. Thus, up to quadratic order,
\begin{equation}
    F \simeq P(\mu) + \frac{\rho_n}{2}\,G_{AB}W^AW^B + 2\lambda\,\delta\mu\,G^{AB}k_{AB} - \frac12\,C^{ABCD}k_{AB}k_{CD},
    \label{eq: SM F second order}
\end{equation}
with $P'(\mu_0)=n$ the number density, $P''(\mu_0)=\chi$ the compressibility, $\rho_n$ the normal density, $\lambda$ the density-strain coupling, and
\begin{equation}
    C^{ABCD} = B\,G^{AB}G^{CD} + 2\mu_e\Big(G^{A(C}G^{D)B} - \frac{1}{d}G^{AB}G^{CD}\Big),
    \label{eq: SM elastic tensor}
\end{equation}
with bulk modulus $B$ and shear modulus $\mu_e$. In $d=2$ this is the parametrization of curved elasticity in \cite{Zhou2025CurvedOdd}. Since $C^{ABCD}$ is a second derivative of $F$, it is symmetric under $(AB)\leftrightarrow(CD)$, so a passive supersolid has no odd elasticity.
 
\subsection{The quadratic action on a curved substrate}
\label{sec: SM quadratic action}

We expand the Goldstones around the equilibrium of Sec.~\ref{sec: SM equilibrium},
\begin{equation}
    \theta=\theta_0+\delta\theta,\qquad X^A=X^A_0+\delta u^A,
\end{equation}
where $\delta\theta$ and each $\delta u^A$ are spacetime scalars, so that $\xi_\mu=\xi_{0\mu}+\partial_\mu\delta\theta$ and $e^A_\mu=e^A_{0\mu}+\partial_\mu\delta u^A$. We write $\dot f\coloneqq v^\mu\partial_\mu f$ and $e_0^{A\mu}\coloneqq h^{\mu\nu}e^A_{0\nu}$.

\textit{Chemical potential and counterflow.} With $\xi_{0\mu}=\mu_0n_\mu$, $v^\mu n_\mu=1$, $h^{\mu\nu}n_\nu=0$ and $v^\mu e^A_{0\mu}=0$, the definitions \eqref{eq: SM variables} give
\begin{equation}
    \delta\mu=\delta\dot\theta-\frac{1}{2\mass}h^{\mu\nu}\partial_\mu\delta\theta\,\partial_\nu\delta\theta,\qquad
    \delta W^A=\frac{1}{\mass} e_0^{A\mu}\partial_\mu\delta\theta-\delta\dot u^A .
    \label{eq: SM dmu dW}
\end{equation}
The first expression is exact. For $W^A$ we keep only the linear part, since $W^A_0=0$ and $W^A$ enters $F$ in pairs; it is the lattice velocity relative to the superfluid, projected on the lattice. The quadratic part of $\delta\mu$ must be kept, because it multiplies the equilibrium density. No advection terms $h^{\mu\nu}\xi_{0\mu}\partial_\nu(\cdot)$ appear, since $h^{\mu\nu}\xi_{0\nu}=0$.

\textit{Strain.} The reference metric is a property of the material, so it is evaluated at the displaced lattice site, and both terms of $k_{AB}=\frac12(G_{AB}-U_{AB})$ vary. Using $U_{0AB}=G_{AB}$ and $e_0^{A\mu}\partial_\mu=G^{AC}\partial_C$, with $\partial_C\equiv\partial/\partial X_0^C$, to linear order
\begin{equation}
    \delta G_{AB}=\delta u^C\partial_CG_{AB},\qquad
    \delta U_{AB}=-G_{AC}\,\partial_B\delta u^C-G_{BC}\,\partial_A\delta u^C ,
\end{equation}
so that
\begin{equation}
    u_{AB}\coloneqq\delta k_{AB}=\frac12\,\mathcal{L}_{\delta u}G_{AB}=\nabla_{(A}\delta u_{B)} ,
    \label{eq: SM fluctuations}
\end{equation}
with $\nabla$ the Levi-Civita connection of $G_{AB}$. The term coming from the explicit dependence $G_{AB}(X)$ is the connection term that makes the strain a tensor: $u_{AB}$ measures how far $\delta u$ is from an isometry of the reference metric. Since $X_0$ is an isometry between the spatial slice and comoving space, Eq.~\eqref{eq: SM stress free}, it maps $\nabla$ to the covariant derivative of the substrate, and with $\delta u_a\coloneqq G_{AB}e^A_{0a}\delta u^B$ the strain is $u_{ab}=\nabla_{(a}\delta u_{b)}$. The flat strain $\partial_{(i}u_{j)}$ thus becomes its covariant version, with no explicit curvature terms. Second-order pieces of $\delta k_{AB}$ multiply $\partial F/\partial k_{AB}|_0$, which vanishes for a stress-free reference state; for a pre-strained material they would add a pre-stress term.

\textit{Quadratic action.} Inserting these fluctuations in \eqref{eq: SM F second order} and using $G_{AB}e_0^{A\mu}e_0^{B\nu}=h^{\mu\nu}$,
\begin{equation}
    \begin{split}
        P(\mu)&\to n\,\delta\dot\theta-\frac{n}{2\mass}h^{\mu\nu}\partial_\mu\delta\theta\,\partial_\nu\delta\theta+\frac{\chi}{2}\delta\dot\theta^{\,2},\\
        \frac{\rho_n}{2}G_{AB}W^AW^B&\to\frac{\rho_n}{2\mass^2}h^{\mu\nu}\partial_\mu\delta\theta\,\partial_\nu\delta\theta-\frac{\rho_n}{\mass}G_{AB}\,\delta\dot u^A e_0^{B\mu}\partial_\mu\delta\theta+\frac{\rho_n}{2}G_{AB}\,\delta\dot u^A\delta\dot u^B,\\
        2\lambda\,\delta\mu\,G^{AB}k_{AB}&\to2\lambda\,\delta\dot\theta\,\Theta,\qquad
        -\frac12C^{ABCD}k_{AB}k_{CD}\to-\frac12C^{ABCD}u_{AB}u_{CD},
    \end{split}
\end{equation}
with $\Theta\coloneqq G^{AB}u_{AB}=\nabla_A\delta u^A$ the dilation of the lattice. The gradient terms of the first two lines combine, with $\rho\coloneqq \mass n$, into $-\rho_s/2\mass^2$ with $\rho_s\coloneqq\rho-\rho_n$. Collecting everything,
\begin{equation}
    \mathcal{L}^{(2)} = \frac{\chi}{2}\,\delta\dot\theta^{\,2} - \frac{\rho_s}{2\mass^2}\,h^{\mu\nu}\partial_\mu\delta\theta\,\partial_\nu\delta\theta + \frac{\rho_n}{2}\,G_{AB}\,\delta\dot u^A\delta\dot u^B - \frac{\rho_n}{\mass}\,G_{AB}\,\delta\dot u^A\,e_0^{B\mu}\partial_\mu\delta\theta
        + 2\lambda\,\delta\dot\theta\,\Theta - \frac12\,C^{ABCD}u_{AB}u_{CD} + n\,\delta\dot\theta ,
    \label{eq: SM L2}
\end{equation}
i.e.\ the compressibility of the condensate, the phase stiffness, the kinetic energy of the lattice, the coupling between lattice motion and superflow, the density--strain coupling and the elastic energy. The last term is a total time derivative on a static background and only carries the equilibrium density into the currents. There is no term linear in $\delta u^A$, because the equilibrium is stress free.

The superfluid stiffness was not put in by hand: it receives contributions from the quadratic part of the Galilean invariant $\mu$ and from the counterflow term, whose sum fixes $\rho_s=\rho-\rho_n$. The two-fluid decomposition of supersolids \cite{Andreev1969} thus follows from Galilean invariance, here and in the flat theory of Sec.~\ref{sec: SM flat benchmark}. The density is not an independent field; from $\rho=\mass\,\partial F/\partial\mu$,
\begin{equation}
    \delta\rho=\mass\chi\,\delta\dot\theta+2\mass\lambda\,\Theta ,
    \label{eq: SM drho}
\end{equation}
so it changes when the chemical potential changes or when the lattice dilates.

On a static substrate, with $h^{\mu\nu}\to\ovg^{ab}$ and indices identified through $X_0^A$, $G_{AB}\delta\dot u^A\delta\dot u^B\to\delta\dot u^a\delta\dot u_a$ and $e_0^{B\mu}\partial_\mu\to\nabla^b$, and \eqref{eq: SM L2} becomes the form used in the main text. Curvature does not appear explicitly: it enters through the covariant derivative in the strain and, in the equations of motion, through the Ricci tensor when covariant derivatives are commuted.
 
\subsection{Flat-space benchmark}
\label{sec: SM flat benchmark}
 
On flat space, \eqref{eq: SM L2} reproduces the supersolid hydrodynamics of Yoo and Dorsey \cite{yoo_hydrodynamic_2010}, with $\delta u^A=-u^A_{\rm YD}$, $\delta\theta=-\mass\varphi$, so that $W^A\to v_n-v_s$. Their Lagrangian has one more field, the density fluctuation, which enters without time derivatives. Integrating it out, the two Lagrangians agree term by term provided
\begin{equation}
    \chi = n_0^2\kappa, \qquad \lambda = \frac{\rho_0^2\kappa\gamma}{2\mass}, \qquad \mu_e = \tilde\mu, \qquad B = \tilde\lambda + \frac{2\tilde\mu}{d} - \rho_0^2\kappa\gamma^2 ,
    \label{eq: SM YD dictionary}
\end{equation}
with $\kappa$, $\gamma$, $\tilde\lambda$ and $\tilde\mu$ their compressibility, density-strain coupling and Lam\'e coefficients at fixed density. Our $B$ is the bulk modulus at fixed chemical potential. It is smaller, since under a slow compression the superfluid can flow out of the unit cell.
 
\subsection{Equations of motion and currents}
\label{sec: SM EOM}
 
Varying \eqref{eq: SM L2} gives two conservation laws,
\begin{equation}
    \Xi_\mu J_\theta^\mu = 0, \qquad \Xi_\mu J_A^\mu = 0, \qquad J_\theta^\mu\coloneqq\frac{\partial\mathcal{L}}{\partial(\partial_\mu\delta\theta)}, \qquad J_A^\mu\coloneqq\frac{\partial\mathcal{L}}{\partial(\partial_\mu\delta u^A)},
    \label{eq: SM EOM}
\end{equation}
where, on the lattice index, the divergence contains the Christoffel symbols $\Gamma^C_{AB}$ of $G_{AB}$,
\begin{equation}
    \Xi_\mu J_A^\mu=\frac{1}{\sqrt{\gm}}\partial_\mu\big(\sqrt{\gm}\,J_A^\mu\big)-\Gamma^C_{AB}\,e^B_{0\mu}J_C^\mu .
    \label{eq: SM covariant divergence}
\end{equation}
The second term comes from the connection term in \eqref{eq: SM fluctuations}. The currents are
\begin{equation}
    \mass\,J_\theta^\mu = (\rho_0+\delta\rho)\,v^\mu + \rho_s\,\delta u_s^\mu + \rho_n\,\delta u_X^\mu, \qquad
    J_A^\mu = -\big(p_A\,v^\mu + \sigma_{AB}\,e_0^{B\mu}\big),
    \label{eq: SM currents}
\end{equation}
with $\delta\rho = \mass\chi\,\delta\dot\theta + 2\mass\lambda\,\Theta$, $\sigma_{AB} = C_{ABCD}\,u^{CD} - 2\lambda\,\delta\dot\theta\,G_{AB}$, the counterflow momentum $p_A=\rho_n G_{AB}\delta W^B$, and the linearized velocities $\delta u_s^\mu=-h^{\mu\nu}\partial_\nu\delta\theta/\mass$ and $\delta u_X^\mu=-e^\mu_{0A}\delta\dot u^A$. The first equation is mass conservation, with the two-fluid flux. The second is the force balance on the lattice, $\partial_t p_A = -\nabla^B\sigma_{BA}$. Together with the Josephson relation $\partial_t\delta u_s=-\nabla\delta\mu/\mass$, it gives momentum conservation with the stress of \eqref{eq: SM improved stress constitutive}.
 
\subsection{Consistency with the Ward identities}
\label{sec: SM EOM Ward}
 
Comparing the variation of the action with \eqref{eq: SM dW supersolid},
\begin{equation}
    U = -\,\Xi_\mu J_\theta^\mu, \qquad K_A = -\,\Xi_\mu J_A^\mu ,
\end{equation}
so the sources of the Ward identities are the Euler-Lagrange operators, and on shell the Ward identities become conservation laws. Moreover, $\theta$ and $A_\mu$ enter only through $\xi_\mu$, so $J_\theta^\mu=\mcJ^\mu/\mass$ exactly, and $X^A$ enters through $e^A_\mu$ in $W^A$ and $U^{AB}$, so $J_A^\mu\coloneqq\partial F/\partial(\partial_\mu X^A)=-\big(p_A u_s^\mu+\sigma_{AB}e^{B\mu}\big)$ to all orders. $X^A$ also enters explicitly through $G_{AB}(X)$. Since $F$ is invariant under reparametrizations of $X^A$, this explicit dependence combines with $\partial_\mu J_A^\mu$ into the covariant divergence \eqref{eq: SM covariant divergence}, with $e^A_\mu$ in place of $e^A_{0\mu}$, so $K_A=-\Xi_\mu J_A^\mu$ also holds to all orders. Eq.~\eqref{eq: SM currents} is the linearization of these, and the currents of the action are the constitutive relations of Sec.~\ref{sec: SM CR}.

\section{Collective modes}
\label{sec: SM collective}

We first write the linear equations on a general surface, then solve them exactly on the sphere, and finally treat the spheroid to first order in its deviation from the sphere.

\subsection{Equations of motion on a surface}
\label{sec: SM sd EOM}

We take a static substrate, $n_\mu dx^\mu=dt$, $v^\mu\partial_\mu=\partial_t$, $A_\mu=0$ and $h^{\mu\nu}\partial_\mu\otimes\partial_\nu=\ovg^{ab}\partial_a\otimes\partial_b$, and keep the material assumptions of Sec.~\ref{sec: SM equilibrium}. Thus, the reference metric is the induced one, $X_0^A$ is an isometry and the strain is $u_{ab}=\nabla_{(a}\delta u_{b)}$. Without a trap all coefficients are constant. Inserting the currents \eqref{eq: SM currents} in \eqref{eq: SM EOM},
\begin{equation}
    \chi\,\delta\ddot\theta+g\,\dot\Theta-\frac{\rho_s}{\mass^2}\Delta\delta\theta=0,\qquad
    \rho_n\,\delta\ddot u_a+g\,\nabla_a\delta\dot\theta-\nabla^b\sigma^{\rm el}_{ba}=0,\qquad g\coloneqq2\lambda-\frac{\rho_n}{\mass},
    \label{eq: SM sd general eom}
\end{equation}
with $\sigma^{\rm el}_{ab}=C_{abcd}u^{cd}$. In two dimensions $R_{ab}=K\ovg_{ab}$, also when $K$ varies, so commuting the derivatives in the elastic force gives, as in curved elasticity \cite{Zhou2025CurvedOdd},
\begin{equation}
    \nabla^b\sigma^{\rm el}_{ba}=B\,\nabla_a\Theta+\mu_e\,\Delta\delta u_a+\mu_e\,K\,\delta u_a .
\end{equation}
We split the displacement into compression and shear, $\delta u_a=\nabla_a\Phi+\epsilon_{ab}\nabla^b\Psi$, so that $\Theta=\Delta\Phi$. Taking the divergence and the curl of the lattice equation,
\begin{equation}
    \begin{split}
        &\chi\,\delta\ddot\theta+g\,\Delta\dot\Phi-\frac{\rho_s}{\mass^2}\Delta\delta\theta=0,\\
        &\Delta\Big[\rho_n\ddot\Phi+g\,\delta\dot\theta-(B+\mu_e)\Delta\Phi\Big]-2\mu_e\nabla_a\big(K\nabla^a\Phi\big)-2\mu_e\,\epsilon^{ab}\nabla_aK\,\nabla_b\Psi=0,\\
        &\Delta\Big[\rho_n\ddot\Psi-\mu_e\Delta\Psi\Big]-2\mu_e\nabla_a\big(K\nabla^a\Psi\big)+2\mu_e\,\epsilon^{ab}\nabla_aK\,\nabla_b\Phi=0 .
    \end{split}
    \label{eq: SM sd eom}
\end{equation}
These equations are exact on any surface with the topology of the sphere. The phase couples only to the compression $\Phi$. Compression and shear couple only through $\epsilon^{ab}\nabla_aK\nabla_b$, which vanishes if and only if $K$ is constant.

\subsection{The sphere}
\label{sec: SM sphere}

On a sphere of radius $R$, $K=1/R^2$, so $\nabla_a(K\nabla^a f)=K\Delta f$ and the coupling between compression and shear vanishes. Every term in the $\Phi$ and $\Psi$ equations of \eqref{eq: SM sd eom} is then a Laplacian, which is invertible on non-constant functions, and
\begin{equation}
    \rho_n\ddot\Phi+g\,\delta\dot\theta-(B+\mu_e)\Delta\Phi-\frac{2\mu_e}{R^2}\Phi=0,\qquad
    \rho_n\ddot\Psi-\mu_e\Delta\Psi-\frac{2\mu_e}{R^2}\Psi=0 .
    \label{eq: SM sphere eom}
\end{equation}
The curvature lowers the elastic stiffnesses, $\mu_e\Lambda_l$ for shear and $(B+\mu_e)\Lambda_l$ for compression, by $2\mu_e/R^2$. The coefficients are constant, so different harmonics do not mix, and we take every field proportional to $Y_{lm}e^{-i\omega t}$, with $\Delta Y_{lm}=-\Lambda_lY_{lm}$ and $\Lambda_l\coloneqq l(l+1)/R^2$. Each level is $(2l+1)$-fold degenerate in $m$.

The shear modes are
\begin{equation}
    \rho_n\,\omega_T^2=\mu_e\Big(\Lambda_l-\frac{2}{R^2}\Big)=\frac{\mu_e}{R^2}\,(l-1)(l+2).
    \label{eq: SM sphere transverse}
\end{equation}
Only the normal component oscillates, and the shear carries no phase and no density. For $l=1$, $\epsilon_{ab}\nabla^bY_{1m}$ are the Killing vectors of the sphere, i.e.\ its three rigid rotations, so $u_{ab}=\nabla_{(a}\delta u_{b)}=0$ and these modes have zero frequency.

For the longitudinal sector, the phase and $\Phi$ equations give
\begin{equation}
    \Big(\chi\omega^2-\frac{\rho_s\Lambda_l}{\mass^2}\Big)\hat\theta=i\omega g\Lambda_l\,\hat\Phi,\qquad
    \big(\rho_n\omega^2-\mcE_l\big)\hat\Phi=-i\omega g\,\hat\theta,\qquad
    \mcE_l\coloneqq(B+\mu_e)\Lambda_l-\frac{2\mu_e}{R^2},
    \label{eq: SM sphere longitudinal rows}
\end{equation}
and the frequencies are the roots of
\begin{equation}
    \big(\omega^2-\omega_\theta^2\big)\big(\omega^2-\omega_E^2\big)=\frac{g^2\Lambda_l}{\chi\rho_n}\,\omega^2,\qquad
    \omega_\theta^2\coloneqq\frac{\rho_s\Lambda_l}{\mass^2\chi},\qquad \omega_E^2\coloneqq\frac{\mcE_l}{\rho_n}.
    \label{eq: SM sphere longitudinal}
\end{equation}
These are the phase mode of the condensate and the compressional mode of the lattice, mixed by $g$. With $x=\omega^2$ the left-hand side minus the right-hand side is an upward parabola, which is $\le0$ at $x=\omega_\theta^2$ and at $x=\omega_E^2$. Thus, the two roots $\omega_\pm^2$ are real and positive, and they enclose both uncoupled frequencies, $\omega_-^2\le\min(\omega_\theta^2,\omega_E^2)$ and $\omega_+^2\ge\max(\omega_\theta^2,\omega_E^2)$. We measure the content of each branch by the elastic and phase weights
\begin{equation}
    Z_\sigma\coloneqq\frac{\omega_\sigma^2-\omega_\theta^2}{\omega_\sigma^2-\omega_{\bar\sigma}^2},\qquad
    Y_\sigma\coloneqq\frac{\omega_\sigma^2-\omega_E^2}{\omega_\sigma^2-\omega_{\bar\sigma}^2},\qquad \sigma\in\{+,-\},
    \label{eq: SM sd weights}
\end{equation}
with $\bar\sigma$ the other branch, $0\le Z_\sigma,Y_\sigma\le1$ and $Z_++Z_-=Y_++Y_-=1$. At $g=0$ the branches are pure. As $\rho_s\to0$, $\omega_-\to0$, so the lower branch is the one that exists because of the superfluid. For $l\gg1$, with $k^2=\Lambda_l$, the curvature terms drop out and the three branches become the transverse sound and the two longitudinal sounds of the flat supersolid \cite{yoo_hydrodynamic_2010}.

\subsection{The spheroid}
\label{sec: SM sd assumptions}

The substrate is a spheroid with equatorial radius $a$ and polar semi-axis $c$, with metric
\begin{equation}
    ds^2=\big(a^2\cos^2\vartheta+c^2\sin^2\vartheta\big)\,d\vartheta^2+a^2\sin^2\vartheta\,d\phi^2,\qquad c^2=a^2(1+\varepsilon).
\end{equation}
$\varepsilon>0$ is prolate, $\varepsilon<0$ oblate and $\varepsilon=0$ the sphere. To first order the Gaussian curvature is
\begin{equation}
    K=\frac{1}{a^2}\big(1+\varepsilon\cos2\vartheta\big),\qquad \cos2\vartheta=\frac13\big(4P_2(\cos\vartheta)-1\big),
    \label{eq: SM sd K}
\end{equation}
a monopole plus a quadrupole. The material is as in Sec.~\ref{sec: SM sd EOM}, so the curvature is the only source of non-uniformity. The spheroid has two symmetries: rotations about its axis, so the azimuthal number $m$ is conserved, and the reflection $z\to-z$. Now $\epsilon^{ab}\nabla_aK\nabla_b\propto\partial_\vartheta K\,\partial_\phi$, so the coupling between compression and shear is proportional to $m$, and the $m=0$ sector decouples exactly.

To first order we use conformal coordinates $(\tilde\vartheta,\phi)$, where the metric is $e^{2\sigma}$ times the round metric of radius $a$, with $e^{2\sigma}=1+\varepsilon\cos^2\vartheta$. The operators in \eqref{eq: SM sd eom} are then the ones of the sphere, corrected at $O(\varepsilon)$ by $e^{\pm2\sigma}$ and by $K$.
 
\subsection{Ansatz and angular integrals}
\label{sec: SM sd ansatz}
 
Time and $\phi$ separate exactly, and we expand the $\vartheta$ dependence around the sphere,
\begin{equation}
    F=\Big(\hat F\,Y_{lm}(\tilde\vartheta,\phi)+\varepsilon\,F^{(1)}\Big)e^{-i\omega t},\qquad \langle Y_{lm}|F^{(1)}\rangle=0,\qquad F\in\{\delta\theta,\Phi,\Psi\}.
\end{equation}
Projecting \eqref{eq: SM sd eom} onto $Y_{lm}$, the correction $F^{(1)}$ drops out of the frequency at first order, provided no level coupled to $(l,m)$ is degenerate with it. The frequencies then follow from the matrix elements of the $O(\varepsilon)$ operators between spherical harmonics. With $L\coloneqq l(l+1)$ and $\Lambda\coloneqq L/a^2$, we need
\begin{equation}
    \begin{split}
        q_{lm}&\coloneqq\langle Y_{lm}|\cos^2\vartheta|Y_{lm}\rangle=\frac{2L-1-2m^2}{(2l-1)(2l+3)},\qquad
        J_{lm}\coloneqq\langle Y_{l+1,m}|\cos\vartheta|Y_{lm}\rangle=\sqrt{\frac{(l+1)^2-m^2}{(2l+1)(2l+3)}},\\
        J^{(2)}_{lm}&\coloneqq\langle Y_{lm}|\cos^2\vartheta|Y_{l+2,m}\rangle=J_{lm}J_{l+1,m},
    \end{split}
    \label{eq:SMsdintegrals}
\end{equation}
The first two are standard \cite{integral_table}, and $J^{(2)}_{lm}$ follows from applying the recursion $\cos\vartheta\,Y_{lm}=J_{lm}Y_{l+1,m}+J_{l-1,m}Y_{l-1,m}$ twice. The gradient elements follow from integrating by parts,
\begin{equation}
    a^2\langle\hat\nabla Y_{lm}|\cos2\vartheta|\hat\nabla Y_{lm}\rangle=2(L-3)\,q_{lm}+2-L,\qquad
    a^2\langle\hat\nabla Y_{lm}|\cos2\vartheta|\hat\nabla Y_{l+2,m}\rangle=2l(l+3)\,J^{(2)}_{lm}.
\end{equation}
The deformation enters the multiplets through the quadrupole weight
\begin{equation}
    p_{lm}\coloneqq\langle Y_{lm}|P_2(\cos\vartheta)|Y_{lm}\rangle=\frac{L-3m^2}{(2l-1)(2l+3)} .
\end{equation}
 
\subsection{Modes of the spheroid}
\label{sec: SM sd modes}
 
Away from crossings, the shear modes are
\begin{equation}
    \rho_n\omega_T^2=\frac{\mu_e}{a^2}(L-2)\Big[1+\frac{\varepsilon}{L}\big(2-q_{lm}(L+6)\big)\Big],
    \label{eq: SM sd transverse}
\end{equation}
and the longitudinal modes are the roots of
\begin{equation}
    \omega^4-\mfS_{lm}\,\omega^2+\mfP_{lm}=0,\qquad
    \mfS_{lm}=\frac{\tilde\mcE_{lm}}{\rho_n}+\Big(\frac{\rho_s}{\mass^2\chi}+\frac{g^2}{\chi\rho_n}\Big)\Lambda'_{lm},\qquad
    \mfP_{lm}=\frac{\rho_s\,\Lambda'_{lm}\,\tilde\mcE_{lm}}{\mass^2\chi\rho_n},
    \label{eq: SM sd longitudinal}
\end{equation}
with $\Lambda'_{lm}=\Lambda(1-\varepsilon q_{lm})$ the eigenvalue of the spheroid Laplacian, and the compressional stiffness
\begin{equation}
    \tilde\mcE_{lm}=(B+\mu_e)\Lambda-\frac{2\mu_e}{a^2}+\frac{\varepsilon\mu_e}{a^2}\,\frac{(L-2)\big(2-q_{lm}(L+6)\big)}{L}-\frac{\varepsilon B}{a^2}\,L\,q_{lm} .
\end{equation}
At $\varepsilon=0$ these are the modes of Sec.~\ref{sec: SM sphere}, and for large $l$ they become the sounds of a flat supersolid \cite{yoo_hydrodynamic_2010}.
 
To isolate the effect of the non-uniform curvature, we compare the spheroid with the sphere of the same area. With $e^{2\sigma}=1+\varepsilon\cos^2\vartheta$,
\begin{equation}
    A=\int\!d\vartheta\,d\phi\,\sqrt{\hat g}\,e^{2\sigma}=4\pi a^2\Big(1+\frac{\varepsilon}{3}\Big),\qquad A\eqqcolon4\pi R^2\quad\Longrightarrow\quad \frac{1}{a^2}=\frac{1}{R^2}\Big(1+\frac{\varepsilon}{3}\Big).
\end{equation}
By Gauss-Bonnet average, the Gaussian curvature is $4\pi/A=1/R^2$, the one of the sphere of radius $R$, so at fixed $R$ the two surfaces differ only by the quadrupole of $K$. Every frequency of Sec.~\ref{sec: SM sd modes} is $a^{-2}$ times a function of $\varepsilon$, so at fixed $R$ each $\omega^2$ picks up the factor $1+\varepsilon/3$. We also write $q_{lm}=\frac13+\frac23p_{lm}$, which isolates the quadrupole. For the shear modes, \eqref{eq: SM sd transverse} becomes
\begin{equation}
    \rho_n\omega_T^2=\frac{\mu_e}{R^2}(L-2)\Big[1+\varepsilon\Big(\frac13+\frac{2-q_{lm}(L+6)}{L}\Big)\Big]=\frac{\mu_e}{R^2}(L-2)\Big[1-\varepsilon\Big(\frac23+\frac4L\Big)p_{lm}\Big],
\end{equation}
where the monopole part of $q_{lm}$ cancels the area factor. For the longitudinal modes, with $\Lambda_R\coloneqq L/R^2$, the same steps give
\begin{equation}
    \Lambda'_{lm}=\Lambda_R\Big(1-\frac23\varepsilon p_{lm}\Big),\qquad
    \tilde\mcE_{lm}=\Big(1-\frac23\varepsilon p_{lm}\Big)\mcE_l-\frac4L\,\varepsilon p_{lm}\,\rho_n\omega_T^2(l),
\end{equation}
with $\mcE_l$ and $\omega_T(l)$ the values on the sphere of radius $R$. The longitudinal frequencies are the zeros of
\begin{equation}
    D(x)=(x-\alpha)(x-e)-c\,x,\qquad \alpha=\frac{\rho_s\Lambda'_{lm}}{\mass^2\chi},\qquad e=\frac{\tilde\mcE_{lm}}{\rho_n},\qquad c=\frac{g^2\Lambda'_{lm}}{\chi\rho_n},
\end{equation}
since $D=x^2-\mfS x+\mfP$. $D$ is homogeneous, so rescaling $(\alpha,e,c)$ by a common factor rescales the roots by the same factor. The factor $1-\frac23\varepsilon p_{lm}$ is common to $\alpha$, $c$ and $e$, and $e$ is shifted in addition by $-\frac4L\varepsilon p_{lm}\,\omega_T^2(l)$. Therefore,
\begin{equation}
    \omega_\sigma^2(\varepsilon)=\Big(1-\frac23\varepsilon p_{lm}\Big)\omega_\sigma^2-\frac4L\,\varepsilon p_{lm}\,\omega_T^2(l)\,\frac{\partial\omega_\sigma^2}{\partial e},\qquad
    \frac{\partial\omega_\sigma^2}{\partial e}=\frac{\omega_\sigma^2-\omega_\theta^2}{\omega_\sigma^2-\omega_{\bar\sigma}^2}=Z_\sigma,
\end{equation}
where the derivative follows from differentiating $D(\omega_\sigma^2)=0$ with respect to $e$. It is the elastic weight of \eqref{eq: SM sd weights}. Collecting, the three branches split as
\begin{equation}
    \omega_s^2(\varepsilon)=\omega_s^2\Big[1-\varepsilon\,\kappa_s(l)\,p_{lm}\Big],\qquad
    \kappa_s(l)=\frac23+\frac4L\,Z_s\,\frac{\omega_T^2(l)}{\omega_s^2},\qquad s\in\{T,+,-\},\quad Z_T\coloneqq1,
    \label{eq: SM sd splitting}
\end{equation}
with all quantities on the right evaluated on the sphere of radius $R$. The $m$-degeneracy is lifted through $p_{lm}$ alone, so the $m$-dependence is fixed by geometry and is the same for all branches. It has the form of the ellipticity splitting of the free oscillations of the Earth \cite{Dahlen1968}, and geometry lifts the degeneracies in the same way for Bose gases on ellipsoids \cite{Tononi2025AxiallySymmetric}. The material enters only through $\kappa_s$: the $\frac23$ is the shift of the spectrum of the Laplacian, common to all branches, and $\frac4L Z_s$ comes from the Ricci term, which acts on the shear part of the stiffness with the elastic weight of each branch. Therefore, the shear multiplets measure $\varepsilon$ alone.

\subsection{Degeneracies and mixing}
\label{sec: SM sd mixing}
 
On the sphere every level is $(2l+1)$-fold degenerate in $m$, and levels with different $l$ can coincide for particular values of the material parameters. The deformation conserves $m$, so it never couples the states of a multiplet, and it only splits them, as in \eqref{eq: SM sd splitting}. The coincidences between different $l$, which we call crossings, are different. Near them the expansion of Sec.~\ref{sec: SM sd ansatz} fails, we have to keep both harmonics in the ansatz, and the two levels mix.
 
Let us first collect the couplings. We project the equations \eqref{eq: SM sd eom} at frequency $\omega$ onto $Y_{lm}$, without dividing the rows. The diagonal entries are then the ones of the sphere, corrected as in Sec.~\ref{sec: SM sd modes},
\begin{equation}
    \begin{split}
    &\mathsf{M}^{(l)}_{\theta\theta}=\chi(1+\varepsilon q_{lm})\,\omega^2-\frac{\rho_s\Lambda_l}{\mass^2},\qquad
    \mathsf{M}^{(l)}_{\theta\Phi}=\overline{\mathsf{M}^{(l)}_{\Phi\theta}}=-i\omega g\Lambda_l,\\
    &\mathsf{M}^{(l)}_{\Phi\Phi}=\Lambda_l\big(\rho_n\omega^2-\tilde\mcE_{lm}\big),\qquad
    \mathsf{M}^{(l)}_{\Psi\Psi}=\Lambda_l\rho_n\big(\omega^2-\omega_T^2(l)\big).
    \end{split}
    \label{eq: SM sd diagonal}
\end{equation}
The off-diagonal entries come from the $O(\varepsilon)$ operators. In the $\Phi$ equation the coupling to the shear is $-2\mu_e\,\epsilon^{ab}\nabla_aK\nabla_b\Psi$, which to first order is $(8\varepsilon\mu_e/a^4)\cos\vartheta\,\partial_\phi\Psi$, and in the $\Psi$ equation the coupling to $\Phi$ has the opposite sign. Since $\cos\vartheta$ connects $l$ to $l\pm1$,
\begin{equation}
    \langle Y_{km}|\cdot|Y_{nm}\rangle_{\Phi\Psi}=iV_{k,n},\quad
    \langle Y_{nm}|\cdot|Y_{km}\rangle_{\Psi\Phi}=-iV_{k,n},\quad
    V_{k,n}\coloneqq\frac{8\varepsilon\mu_e m}{a^4}\,\langle Y_{km}|\cos\vartheta|Y_{nm}\rangle,
    \label{eq: SM sd V}
\end{equation}
which is real and symmetric, and nonzero only for $|k-n|=1$. The conformal factor and $K-a^{-2}=\varepsilon\cos2\vartheta/a^2$ connect $l$ to $l$ and $l\pm2$, and between $Y_{lm}$ and $Y_{l+2,m}$ they give
\begin{equation}
    C_{\theta\theta}=\varepsilon\chi\omega^2J^{(2)}_{lm},\qquad
    C_{\Phi\Phi}=\frac{\varepsilon J^{(2)}_{lm}}{a^4}\Big[(B+\mu_e)L_lL_{l+2}+4\mu_e\,l(l+3)\Big],\qquad
    C_{\Psi\Psi}=\frac{\varepsilon\mu_eJ^{(2)}_{lm}}{a^4}\Big[L_lL_{l+2}+4\,l(l+3)\Big].
    \label{eq: SM sd C}
\end{equation}
In $C_{\Phi\Phi}$ and $C_{\Psi\Psi}$ the first term comes from the conformal factor and the second from the Ricci term. The conformal factor and $K$ are even under $z\to-z$, while $\cos\vartheta\,\partial_\phi$ is odd, and this gives the selection rules:
\begin{enumerate}
    \item $\Delta l=\pm1$ couples only compression and shear, $\Phi\leftrightarrow\Psi$, through the gradient of $K$, with strength $V\propto\varepsilon m J_{lm}$. It vanishes for $m=0$;
    \item $\Delta l=\pm2$ couples each field with itself, $\delta\theta\leftrightarrow\delta\theta$, $\Phi\leftrightarrow\Phi$, $\Psi\leftrightarrow\Psi$, with strength $\propto\varepsilon J^{(2)}_{lm}$. It survives for $m=0$;
    \item $|\Delta l|\ge3$ does not couple at first order.
\end{enumerate}
 
Now the crossings. Levels of the same type never cross, since $\omega_T$, $\omega_\theta$ and $\omega_E$ grow with $l$. Across $\Delta l=\pm1$, a longitudinal level of $l$ crosses the shear level of $l\pm1$. At $g=0$, for instance,
\begin{equation}
    \omega_E(l)=\omega_T(l+1)\iff\frac{B}{\mu_e}=\frac{2}{l},\qquad
    \omega_\theta(l)=\omega_T(l+1)\iff\frac{\rho_s\rho_n}{\mass^2\chi\mu_e}=\frac{l+3}{l+1},
\end{equation}
and for $g\neq0$ these points move continuously. The pair is coupled only if $m\neq0$. Across $\Delta l=\pm2$, the only coincidences are $\omega_E(l)=\omega_\theta(l+2)$ and $\omega_\theta(l)=\omega_E(l+2)$, i.e.\ $\rho_nc_\theta^2L_{l+2}=(B+\mu_e)L_l-2\mu_e$ and the same with $L_l\leftrightarrow L_{l+2}$, where $c_\theta^2\coloneqq\rho_s/(\mass^2\chi)$. Both are crossings $\omega_+(l)=\omega_-(l+2)$, and both pair the phase mode of one multiplet with the elastic mode of the other. Therefore, they exist only in a supersolid.
 
\emph{Mixing across $\Delta l=\pm1$.} Let us take $\omega_\sigma(l)\simeq\omega_T(l+1)$, $\sigma\in\{+,-\}$, and keep $(\delta\theta,\Phi)$ in $Y_{lm}$ and $\Psi$ in $Y_{l+1,m}$. The amplitudes obey
\begin{equation}
    \begin{pmatrix}
        \mathsf{M}^{(l)}_{\theta\theta} & -i\omega g\Lambda_l & 0\\
        i\omega g\Lambda_l & \Lambda_l\big(\rho_n\omega^2-\tilde\mcE_{lm}\big) & iV\\
        0 & -iV & \Lambda_{l+1}\rho_n\big(\omega^2-\omega_T^2(l+1)\big)
    \end{pmatrix}
    \begin{pmatrix}\hat\theta\\ \hat\Phi\\ \hat\Psi\end{pmatrix}=0,\qquad V\equiv V_{l,l+1}.
    \label{eq: SM sd 3x3}
\end{equation}
The upper $2\times2$ block is the longitudinal problem of $l$, with determinant $\propto(\omega^2-\omega_+^2)(\omega^2-\omega_-^2)$. Therefore, the determinant of \eqref{eq: SM sd 3x3} gives
\begin{equation}
    \big(\omega^2-\omega_T^2\big)\big(\omega^2-\omega_+^2\big)\big(\omega^2-\omega_-^2\big)=\frac{\hat V^2}{\rho_n^2}\big(\omega^2-\omega_\theta'^2\big),\qquad
    \hat V\coloneqq\frac{V}{\sqrt{\Lambda_l\Lambda_{l+1}}},
    \label{eq: SM sd cubic}
\end{equation}
with $\omega_T\equiv\omega_T(l+1)$ and $\omega_\theta'^2\coloneqq\rho_s\Lambda'_{lm}/(\mass^2\chi)$. Near the crossing the other branch $\bar\sigma$ is far, and \eqref{eq: SM sd cubic} reduces to the two-level problem
\begin{equation}
    \big(\omega^2-\omega_\sigma^2\big)\big(\omega^2-\omega_T^2\big)=|U|^2,\qquad U=\frac{\hat V}{\rho_n}\sqrt{Z_\sigma},\qquad
    \Delta(\omega^2)=\sqrt{\big(\omega_\sigma^2-\omega_T^2\big)^2+4|U|^2},
    \label{eq: SM sd two level}
\end{equation}
valid for $|U|\ll|\omega_\sigma^2-\omega_{\bar\sigma}^2|$. The levels repel, and the minimal gap is
\begin{equation}
    \Delta(\omega^2)_{\min}=2|U|=\frac{16|\varepsilon|\mu_e}{\rho_na^2}\,\frac{|m|\,J_{lm}}{\sqrt{L_lL_{l+1}}}\,\sqrt{Z_\sigma}.
    \label{eq: SM sd gap 1}
\end{equation}
The shear mode only talks to the elastic content $Z_\sigma$ of the longitudinal level, since the coupling acts on $\Phi$. At $g=0$ the phase mode is pure, $Z=0$, and \eqref{eq: SM sd cubic} factorizes. For small $g$, $Z_-\simeq g^2\Lambda_l\omega_\theta^2/\big(\chi\rho_n(\omega_E^2-\omega_\theta^2)^2\big)$ for $\omega_\theta<\omega_E$, so the phase--shear gap is $\propto|\varepsilon m g|$. The mixed modes follow from the first and third rows:
\begin{equation}
    \hat\theta=\frac{i\omega g\Lambda_l}{\chi\big(\omega^2-\omega_\theta^2\big)}\,\hat\Phi,\qquad
    \hat\Psi=\frac{iV}{\Lambda_{l+1}\rho_n\big(\omega^2-\omega_T^2\big)}\,\hat\Phi,\qquad
    \bigg|\frac{\hat\Psi}{\hat\Phi}\bigg|_{\omega_\sigma=\omega_T}=\sqrt{\frac{L_l}{L_{l+1}}}\,\frac{1}{\sqrt{Z_\sigma}},
    \label{eq: SM sd amplitudes}
\end{equation}
and we color each root in Fig.~\ref{fig:modes} by its shear fraction, $\rho_n\Lambda_{l+1}|\hat\Psi|^2/\big(\chi|\hat\theta|^2+\rho_n\Lambda_l|\hat\Phi|^2+\rho_n\Lambda_{l+1}|\hat\Psi|^2\big)$.

At the crossing both roots are hybrids, with comparable weights of shear and compression, and through $g$ the compression carries a phase. Thus, a shear mode acquires a superflow, which is what we use in Sec.~\ref{sec: SM sd shear}.
 
\emph{Mixing across $\Delta l=\pm2$.} Let us take $\omega_+(l)\simeq\omega_-(l+2)$ and keep $(\delta\theta,\Phi)$ in both harmonics. With $A_k$ the longitudinal $2\times2$ block of $k$ and $C\coloneqq\mathrm{diag}(C_{\theta\theta},C_{\Phi\Phi})$,
\begin{equation}
    \begin{pmatrix}A_l(\omega)&C\\ C&A_{l+2}(\omega)\end{pmatrix}
    \begin{pmatrix}\hat\theta^{(l)}\\ \hat\Phi^{(l)}\\ \hat\theta^{(l+2)}\\ \hat\Phi^{(l+2)}\end{pmatrix}=0 .
\end{equation}
Near the crossing only the pole of $A_l^{-1}$ at $\omega_\sigma^2(l)$ matters, and its residue is $v_\sigma v_\sigma^\dagger$, with
\begin{equation}
    v_\sigma=\begin{pmatrix}\sqrt{Y_\sigma/\chi}\\ -i\,s_\sigma\sqrt{Z_\sigma/(\Lambda_l\rho_n)}\end{pmatrix},\qquad s_\sigma\coloneqq\mathrm{sgn}\big(g(\omega_\sigma^2-\omega_{\bar\sigma}^2)\big),
\end{equation}
the normalized sphere eigenvector of the level, and the same at $l+2$. Thus, we get again the two-level problem \eqref{eq: SM sd two level}, now between $\omega_+(l)$ and $\omega_-(l+2)$, with
\begin{equation}
    U=v_-^\dagger C\,v_+=\frac{C_{\theta\theta}}{\chi}\sqrt{Y_+(l)Y_-(l+2)}-\frac{C_{\Phi\Phi}}{\rho_n\sqrt{\Lambda_l\Lambda_{l+2}}}\sqrt{Z_+(l)Z_-(l+2)} .
\end{equation}
In both cases, the gap is twice the matrix element of the deformation between the two sphere eigenvectors. Here $C$ has two parts, from the conformal factor and from the Ricci term, and at the crossing the first one cancels exactly. Indeed, both levels have the same frequency, so $\Lambda_l$ and $\Lambda_{l+2}$ are the two roots of the same quadratic, and this fixes $Y_+Y_-$ and $Z_+Z_-$ so that the conformal contributions of $C_{\theta\theta}$ and $C_{\Phi\Phi}$ are equal. Only the Ricci term is left, i.e. the term $-2\mu_e\nabla_a\big(K\nabla^a\cdot\big)$,
\begin{equation}
    \Delta(\omega^2)_{\min}=\frac{8|\varepsilon|\mu_e}{\rho_na^2}\,\frac{l(l+3)\,J^{(2)}_{lm}}{\sqrt{L_lL_{l+2}}}\,\sqrt{Z_+(l)Z_-(l+2)} .
    \label{eq: SM sd gap 2}
\end{equation}
For small $g$, at the crossing $\omega_E(l)=\omega_\theta(l+2)\eqqcolon\omega_\times$,
\begin{equation}
    \Delta(\omega^2)_{\min}=\frac{4|\varepsilon g|\,\mu_e}{(B+\mu_e)\,a}\,\frac{l(l+3)\,J^{(2)}_{lm}}{(2l+3)\sqrt{\chi\rho_nL_l}}\,\omega_\times+O(g^3),
\end{equation}
and the same with $L_l\to L_{l+2}$ for $\omega_\theta(l)=\omega_E(l+2)$. This gap needs $g$, since without it the phase and elastic modes do not talk to each other, it survives for $m=0$, and it is proportional to the shear modulus times the non-uniformity of $K$. It is specific to a supersolid on a surface of non-constant curvature.
 
\emph{Mixing along the shape.} So far we tuned a material parameter through a crossing at fixed $\varepsilon$. In an experiment one would rather fix the material and change the shape. By \eqref{eq: SM sd splitting}, at fixed area every level is linear in $\varepsilon$. For the pair $\omega_\sigma(l)$, $\omega_T(l+1)$,
\begin{equation}
    \omega_s^2(\varepsilon)=\omega_s^2+s_s\,\varepsilon,\qquad s_s=-\kappa_s\,p_{lm}\,\omega_s^2,\qquad s\in\{\sigma,T\},
    \label{eq: SM sd slopes}
\end{equation}
with $p_{lm}$ taken at $l$ for $\sigma$ and at $l+1$ for $T$, and everything on the right evaluated on the sphere. Since $Z_s\ge0$, $\kappa_s\ge\frac23>0$, so the sign of each slope is the sign of $-p_{lm}$. The detuning $\delta(\varepsilon)=\omega_\sigma^2(\varepsilon)-\omega_T^2(\varepsilon)=\delta_0+s\,\varepsilon$, with $s=s_\sigma-s_T$, is a straight line that vanishes at $\varepsilon_\times=-\delta_0/s$, and we take the material close to a crossing, $\delta_0=O(\varepsilon)$. The coupling is also linear, $U=u\,\varepsilon$, with $u$ read from \eqref{eq: SM sd gap 1} at $\varepsilon_\times$. The two-level problem \eqref{eq: SM sd two level} then gives
\begin{equation}
    x_\pm(\varepsilon)=\frac{\omega_\sigma^2(\varepsilon)+\omega_T^2(\varepsilon)}{2}\pm\frac12f(\varepsilon),\qquad f=\sqrt{\delta^2+4u^2\varepsilon^2}\ \ge\ 2|U| .
\end{equation}
Far from $\varepsilon_\times$, $f\simeq|\delta|$, so the upper root sits on the higher line and the lower root on the lower one. At $\varepsilon_\times$ the lines exchange order, but the roots cannot cross, since $x_+-x_-=f\ge2|U|$. Thus, the lower root runs along one line, bends near $\varepsilon_\times$, and continues along the other. The content of each mode stays with its line, which is why the shear fraction in Fig.~\ref{fig:modes}(b,f) is exchanged between the two roots across the gap.

If the two slopes have opposite signs, the lower root rises on one side of $\varepsilon_\times$ and falls on the other, so it has a maximum, and the upper root a minimum. More precisely, $dx_\mp/d\varepsilon=\frac12(s_\sigma+s_T)\mp\frac12f'$. Since $f$ is the length of a vector linear in $\varepsilon$, it is convex, and $f'$ grows monotonically from $-\sqrt{s^2+4u^2}$ to $+\sqrt{s^2+4u^2}$. Each derivative then vanishes once if and only if
\begin{equation}
    s_\sigma\,s_T<u^2 ,
    \label{eq: SM sd turning}
\end{equation}
which holds for opposite slopes whenever $u\neq0$. With equal signs the roots turn only if the coupling grows faster than the levels move. The closest approach is at $\varepsilon_{\min}=s^2\varepsilon_\times/(s^2+4u^2)$, with separation $2|u\,\delta_0|/\sqrt{s^2+4u^2}$, and the two turning points lie on either side of it.

Opposite slopes require $p_{lm}\,p_{l+1,m}<0$, i.e.\ $L_l<3m^2<L_{l+1}$, which holds for $(l,|m|)=(1,1),(4,3),(6,4),(8,5),\dots$ and does not depend on the material. For $m=0$ nothing turns, for two independent reasons: $p_{l0}=L_l/[(2l-1)(2l+3)]>0$ for every $l$, so the slopes have equal signs, and $u\propto m$ vanishes, so the two levels cross freely, as the $|m|=0$ pair of Fig.~\ref{fig:modes}(b,f). Thus, for opposite slopes a turning point of a mode frequency as a function of the aspect ratio requires only $u\neq0$, i.e.\ the compression--shear coupling $\epsilon^{ab}\nabla_aK\nabla_b$. The same analysis applies across $\Delta l=\pm2$, with $u$ from \eqref{eq: SM sd gap 2}, which survives at $m=0$, but there turning needs $u^2>s_\sigma s_{\sigma'}$ and depends on the material.
 
\subsection{Superfluid response to shear}
\label{sec: SM sd shear}
 
On the sphere a shear mode moves only the lattice: $\Psi$ is decoupled from $\Phi$ and $\delta\theta$, and carries no superflow and no density. On the spheroid the mixing of Sec.~\ref{sec: SM sd mixing} changes this. We drive the lattice with a pure shear force, $f_a=\epsilon_{ab}\nabla^b\psi_f$, added to the right-hand side of the lattice equation. It has no divergence, so it only enters the $\Psi$ equation of \eqref{eq: SM sd eom}, as the right-hand side $\Delta\psi_f$. We take a single harmonic, $\psi_f=\hat\psi_fY_{Lm}e^{-i\omega t}$.
 
By the selection rules, the shear in $Y_{Lm}$ feeds the compression and the phase in the sectors $k=L\pm1$ (mixing across $\Delta l=\pm1$), and the shear in the partners $L'=L\pm2$ (mixing across $\Delta l=\pm2$). Therefore, we keep
\begin{equation}
    \Psi=\Big(\hat\Psi\,Y_{Lm}+\sum_{L'=L\pm2}\hat\Psi_{L'}Y_{L'm}\Big)e^{-i\omega t},\qquad
    \Phi=\sum_{k=L\pm1}\hat\Phi_kY_{km}\,e^{-i\omega t},\qquad
    \delta\theta=\sum_{k=L\pm1}\hat\theta_kY_{km}\,e^{-i\omega t},
\end{equation}
and project with \eqref{eq: SM sd diagonal}, \eqref{eq: SM sd V} and \eqref{eq: SM sd C}, keeping the leading order of each amplitude. For each sector $k$, the rows are the first two rows of \eqref{eq: SM sd 3x3},
\begin{equation}
    \mathsf{M}^{(k)}_{\theta\theta}\hat\theta_k-i\omega g\Lambda_k\hat\Phi_k=0,\qquad
    i\omega g\Lambda_k\hat\theta_k+\Lambda_k\big(\rho_n\omega^2-\tilde\mcE_{km}\big)\hat\Phi_k+i\mathcal S_k=0,\qquad
    \mathcal S_k\coloneqq V_{k,L}\hat\Psi+V_{k,L'}\hat\Psi_{L'},
    \label{eq: SM sd sector rows}
\end{equation}
where $L'$ is the partner adjacent to $k$. The sector is fed by the shear through the $\Delta l=\pm1$ coupling, directly from $L$ and through the partner. The partners obey the $\Delta l=\pm2$ row, and the drive harmonic collects everything,
\begin{equation}
    \Lambda_{L'}\rho_n\big(\omega^2-\omega_T^2(L')\big)\hat\Psi_{L'}+C_{L'}\hat\Psi=0,\qquad
    \Lambda_L\rho_n\big(\omega^2-\omega_T^2\big)\hat\Psi-i\sum_kV_{k,L}\hat\Phi_k+\sum_{L'}C_{L'}\hat\Psi_{L'}=-\Lambda_L\hat\psi_f,
    \label{eq: SM sd drive rows}
\end{equation}
with $C_{L'}$ the $C_{\Psi\Psi}$ of \eqref{eq: SM sd C} between $L$ and $L'$, and $\omega_T\equiv\omega_T(L)$.
 
Solving \eqref{eq: SM sd sector rows}, with $Q_k(x)\coloneqq\big(x-\omega_{+,k}^2\big)\big(x-\omega_{-,k}^2\big)$ the longitudinal levels of the sector and $\omega_{\theta,k}^2\coloneqq\rho_s\Lambda_k/(\mass^2\chi)$,
\begin{equation}
    \hat\theta_k=\frac{\omega\,g\,\mathcal S_k}{\chi\rho_n\,Q_k(\omega^2)},\qquad
    \hat\Phi_k=-\frac{i\,\mathcal S_k\big(\omega^2-\omega_{\theta,k}^2\big)}{\rho_n\Lambda_k\,Q_k(\omega^2)},\qquad
    \hat\Psi_{L'}=-\frac{C_{L'}\,\hat\Psi}{\Lambda_{L'}\rho_n\big(\omega^2-\omega_T^2(L')\big)} .
    \label{eq: SM sd transfer}
\end{equation}
These relations do not contain the source, so they hold for any shear motion, driven or free. Away from the crossings they are $O(\varepsilon)\hat\Psi$, and they become large when $\omega^2$ approaches a longitudinal level of $k$, i.e.\ at the $\Delta l=\pm1$ crossings. Inserting them in \eqref{eq: SM sd drive rows},
\begin{equation}
    \hat\Psi=-\frac{\hat\psi_f}{\rho_n}\,\frac{1}{\omega^2-\omega_T^2-\Sigma(\omega^2)},\qquad
    \Sigma(x)=\sum_{k=L\pm1}\sum_{\sigma=\pm}\frac{\hat V_k^2}{\rho_n^2}\,\frac{Z_{\sigma,k}}{x-\omega_{\sigma,k}^2}+\sum_{L'=L\pm2}\frac{\hat C_{L'}^2}{\rho_n^2}\,\frac{1}{x-\omega_T^2(L')},
    \label{eq: SM sd propagator}
\end{equation}
with $\hat V_k\coloneqq V_{k,L}/\sqrt{\Lambda_k\Lambda_L}$ and $\hat C_{L'}\coloneqq C_{L'}/\sqrt{\Lambda_{L'}\Lambda_L}$. On the sphere $\Sigma=0$ and the shear has a single pole. On the spheroid, the poles of $\Sigma$ are the levels coupled to the shear mode, and the residues are the squared couplings $|U|^2$ of Sec.~\ref{sec: SM sd mixing}. The two longitudinal poles of each sector recombine, since by partial fractions $\sum_{\sigma}Z_{\sigma,k}/(x-\omega_{\sigma,k}^2)=(x-\omega_{\theta,k}^2)/Q_k(x)$: the elastic weights are the residues of this function. Multiplying $\omega^2-\omega_T^2-\Sigma$ by $Q_{L-1}Q_{L+1}T$, with $T(x)\coloneqq\prod_{L'}\big(x-\omega_T^2(L')\big)$, then clears all denominators,
\begin{equation}
    P(x)\coloneqq\big(x-\omega_T^2\big)Q_{L-1}Q_{L+1}T-\sum_k\frac{\hat V_k^2}{\rho_n^2}\big(x-\omega_{\theta,k}^2\big)Q_{\bar k}T-\sum_{L'}\frac{\hat C_{L'}^2}{\rho_n^2}\,Q_{L-1}Q_{L+1}\frac{T}{x-\omega_T^2(L')},
    \label{eq: SM sd secular}
\end{equation}
where $\bar k$ is the other sector, and $\hat\Psi=-\hat\psi_f\,Q_{L-1}Q_{L+1}T/(\rho_nP)$. $P$ is a polynomial of degree seven, and its roots are the frequencies of the coupled modes: one for the shear, two for each sector and one for each partner. These are the mixed modes of Sec.~\ref{sec: SM sd mixing}, and near a crossing two of them are separated by the gap \eqref{eq: SM sd gap 1}.
 
Now the responses. With \eqref{eq: SM sd transfer} and \eqref{eq: SM sd propagator}, the source of the sector $k$ is
\begin{equation}
    \mathcal S_k=-\frac{\hat\psi_f}{\rho_n}\,\frac{Q_k\,Q_{\bar k}\,\mathcal V_k}{P},\qquad
    \mathcal V_k(x)\coloneqq V_{k,L}\,T(x)-\frac{V_{k,L'}C_{L'}}{\Lambda_{L'}\rho_n}\,\frac{T(x)}{x-\omega_T^2(L')} .
\end{equation}
$\mathcal V_k$ is a polynomial, since $T$ contains the factor $x-\omega_T^2(L')$. Its first term is the direct path, shear in $L$ to compression in $k$ across $\Delta l=\pm1$. The second is the path through the partner, across $\Delta l=\pm2$ and then $\Delta l=\pm1$. Thus, the shear, the superfluid velocity, with amplitude $\sqrt{\Lambda_k}|\hat\theta_k|/\mass$, and the density, $\delta\hat n_k=-i\omega\chi\hat\theta_k-2\lambda\Lambda_k\hat\Phi_k$, respond as
\begin{equation}
    \frac{\hat\Psi}{\hat\psi_f}=-\frac{Q_{L-1}\,Q_{L+1}\,T}{\rho_n\,P},\qquad
    \frac{|v_{s,k}|}{|\hat\psi_f|}=\frac{\sqrt{\Lambda_k}\,\omega\,|g|\,\big|Q_{\bar k}\,\mathcal V_k\big|}{\mass\chi\rho_n^2\,|P|},\qquad
    \frac{\delta\hat n_k}{\hat\psi_f}=-\frac{i\,\big[(\rho_n/\mass)\,\omega^2-2\lambda\,\omega_{\theta,k}^2\big]\,Q_{\bar k}\,\mathcal V_k}{\rho_n^2\,P},
    \label{eq: SM sd responses}
\end{equation}
all evaluated at $x=\omega^2$. In the density we used $2\lambda-g=\rho_n/\mass$.
 
These three equations show how the mixing enters. First, they have the same denominator $P$, so they all resonate at the coupled modes, and only there. In the superflow and the density the factor $Q_k$ of $\mathcal S_k$ cancels the $1/Q_k$ of \eqref{eq: SM sd transfer}, so there is no pole at the uncoupled longitudinal levels: the superfluid responds to a shear drive only at frequencies of modes that contain shear. Second, the superflow and the density are proportional to $\mathcal V_k$, i.e.\ to the couplings $V\propto\varepsilon m$ and $C\propto\varepsilon$. They vanish on the sphere, and also for $m=0$, where the $\Delta l=\pm1$ mixing is absent. The $\Delta l=\pm2$ mixing alone moves weight between shear modes, without superflow. Third, near a $\Delta l=\pm1$ crossing two roots of $P$ are separated by the gap \eqref{eq: SM sd gap 1}, so the region where the superfluid responds strongly has a width $\propto\varepsilon m\sqrt{Z_\sigma}$.
 
The height of the response at the crossing follows from the mode content. For a mode at $\omega^2=x$, the ratio of the superflow to the shear velocity $\omega\sqrt{\Lambda_L}|\hat\Psi|$ follows from \eqref{eq: SM sd transfer},
\begin{equation}
    R_k=\frac{|v_{s,k}|}{|v_{\rm shear}|}=\frac{|g|\,|\hat V_k|\,\Lambda_k}{\mass\chi\rho_n\,|Q_k(x)|}.
\end{equation}
Away from the crossings $R_k\propto\varepsilon m g$ is small. At a crossing $\omega_T(L)=\omega_\sigma(k)$, the roots of the two-level problem \eqref{eq: SM sd two level} have $|Q_k|=|\hat V_k|\sqrt{Z_\sigma}\,|\omega_\sigma^2-\omega_{\bar\sigma}^2|/\rho_n$, and the coupling cancels,
\begin{equation}
    R_\times=\frac{1}{\mass\,\omega_\sigma}\sqrt{\frac{\rho_n\Lambda_kY_\sigma}{\chi}} .
    \label{eq: SM sd R cross}
\end{equation}
$R_\times$ does not depend on $\varepsilon$ or on the azimuthal number $m$: the deformation decides where the crossing is and how wide it is, through the mixing, and the material decides how much superflow the hybrid mode carries. At the phase--shear crossing for small $g$, $Y_-\to1$ and $\omega_-^2\to\omega_\theta^2$, so
\begin{equation}
    R_\times^2\simeq\frac{\rho_n}{\rho_s},\qquad\text{i.e.}\qquad \rho_s|v_s|^2\simeq\rho_n|v_{\rm shear}|^2 ,
\end{equation}
and the kinetic energy of the hybrid mode is shared equally between the superflow and the shear of the lattice, however small $\varepsilon$ is. At the elastic-shear crossing, instead, $Y_+=O(g^2)$ and $R_\times=O(g)$.
 
Finally, the superflow is odd in $\omega$, so a static shear does not move the superfluid. The density instead stays finite,
\begin{equation}
    \frac{\delta\hat n_k}{\hat\psi_f}\bigg|_{\omega=0}=\frac{2i\lambda\,\omega_{\theta,k}^2\,Q_{\bar k}(0)\,\mathcal V_k(0)}{\rho_n^2\,P(0)} .
\end{equation}
A static shear of the spheroid is converted into a static compression through the $\Delta l=\pm1$ mixing, and $\lambda$ turns it into a density. The density also vanishes at $\omega^2=2\mass\lambda\,\omega_{\theta,k}^2/\rho_n$, where the phase and compression contributions cancel. Therefore, a shear drive, invisible to a density or superflow probe on the sphere, becomes visible on the spheroid, and most of all near the crossings, where the mixing makes the shear of the lattice drive the largest superflow.

\section{Experimental values}
\label{sec: experimental values}

The literature does not use our variables. Microscopic calculations give the superfluid fraction $f_s=\rho_s/\rho$ and the elastic parameters of the quadratic Lagrangian of Yoo and Dorsey, in the form of Eq.~(10) of Ref.~\cite{Poli2024Excitations2dSS}: $\alpha_{\rho\rho}$, $\alpha_{\rho u}$, $\alpha_{uu}=\tilde\lambda+2\tilde\mu$ and $\tilde\mu$, with $\rho$ the areal number density. Integrating out the density fluctuation as in Sec.~\ref{sec: SM flat benchmark},
\begin{equation}
    \chi=\frac{1}{\alpha_{\rho\rho}},\qquad 2\lambda=\frac{\alpha_{\rho u}}{\alpha_{\rho\rho}},\qquad \mu_e=\tilde\mu,\qquad B+\mu_e=\alpha_{uu}-\frac{\alpha_{\rho u}^2}{\alpha_{\rho\rho}},\qquad (\rho_s,\rho_n)=\mass\rho\,(f_s,1-f_s).
    \label{eq: SM Poli dictionary}
\end{equation}
In flat space the modes of Sec.~\ref{sec: SM sphere} become sounds, $\omega=ck$. The transverse sound has $c_T^2=\mu_e/\rho_n$, and the longitudinal sounds $c_+>c_-$ are the roots of
\begin{equation}
    \big(c^2-c_\theta^2\big)\big(c^2-c_E^2\big)=\frac{g^2}{\chi\rho_n}\,c^2,\qquad c_\theta^2=\frac{\rho_s}{\mass^2\chi},\qquad c_E^2=\frac{B+\mu_e}{\rho_n},
    \label{eq: SM flat sounds}
\end{equation}
the phase mode and the lattice compression mixed by $g$. As in the two-fluid hydrodynamics of helium, with the lattice in the role of the normal component, $c_+$ is called first sound and $c_-$ second sound. With \eqref{eq: SM Poli dictionary}, the sum and product of $c_\pm^2$ coincide with Eqs.~(11)--(13) of Ref.~\cite{Poli2024Excitations2dSS}, so $c_+=c_1$, $c_-=c_2$ and $c_T=c_t$. In terms of their characteristic speeds $c_\kappa$ and $c_u$, Eqs.~(16)--(17), $c_+c_-=\sqrt{f_s}\,c_\kappa c_u$, and for $\lambda=0$, $c_\theta^2=f_sc_\kappa^2$ and $c_E=c_u$.

The modes depend on the material only through four ratios, which are invariant under a rescaling of $\delta\theta$,
\begin{equation}
    f_s,\qquad \frac{B}{\mu_e}=\frac{\alpha_{uu}-\alpha_{\rho u}^2/\alpha_{\rho\rho}}{\tilde\mu}-1,\qquad \frac{c_\theta^2}{c_T^2}=\frac{f_s(1-f_s)\rho^2\alpha_{\rho\rho}}{\tilde\mu},\qquad \frac{2\lambda}{n}=\frac{\alpha_{\rho u}}{\rho\,\alpha_{\rho\rho}},
    \label{eq: SM code ratios}
\end{equation}
and the code uses units $R=\rho_n=\mu_e=\chi=1$. Frequencies are then in units of $c_T/R$ and are converted with $f=\omega\,c_T/(2\pi R)$, the frequency unit of the tables. The damping enters as $\omega^2\to\omega^2+i\Gamma\omega$, in the same units. For Dy, $c_u\ll c_\kappa$. We fix $B$ from $c_2\simeq\sqrt{f_s}\,c_u$, Eq.~(19) of Ref.~\cite{Poli2024Excitations2dSS}, and neglect $\alpha_{\rho u}$, which shifts the speeds by less than $0.4\%$~\cite{Poli2024Excitations2dSS} and is also neglected for dipolar stripes~\cite{2026arXiv260401751P}. For the soft-core gas we use the elastic parameters of Ref.~\cite{Poli2024Excitations2dSS} at $\Lambda=42.5$ directly.

\begin{table*}[ht!]
\caption{$^{164}$Dy values used in Fig.~\ref{fig:modes}(a)--(d). Code units $R=\rho_n=\mu_e=\chi=1$.}
\label{tab:dy}
\begin{ruledtabular}
\begin{tabular}{l l l}
Quantity & Value & Source and reasoning \\
\colrule
Protocol & evaporation & lifetime $>1$~s~\cite{Sohmen2021Birth}; 152~ms after a quench~\cite{Chomaz2019LongLived} \\
Radius $R$ & $15~\mu$m & full $10~\mu$m shell~\cite{Jia2022ShellExpansion}; $35.8~\mu$m microgravity design~\cite{Veyron2025AllOptical} \\
Spacing $d$ & $1.5~\mu$m ($R/d=10$) & $k^*/2\pi=0.206/l$, so $d=4.85\,l$~\cite{Hertkorn2021Pattern}; checked against $d=3.7~\mu$m in Ref.~\cite{Biagioni2024Josephson} \\
Radial confinement & 645~Hz & $l_r=d/4.85$, $f_r=\hbar/(2\pi M l_r^2)$ \\
$f_s=\rho_s/\rho$ & 0.5 & measured 0.8--0.1~\cite{Biagioni2024Josephson}; computed 0.9--0.15 in 2D~\cite{Poli2024Excitations2dSS} \\
$c_T$ & 0.69~mm/s & 0.23~mm/s at 72.4~Hz~\cite{Poli2024Excitations2dSS,SenarathYapa2025Shear}, scaled as $\sqrt{f_r}$ \\
$c_\theta^2/c_T^2$ & 24 & $f_s c_\kappa^2$, with $c_\kappa\simeq c_1\simeq7c_T$ (bulk-incompressible limit) \\
$\lambda$, $g$ & $0$, $-\rho_n/M$ & density--strain term neglected; it shifts the sound speeds by $<0.4\%$ \\
$B/\mu_e$ & 3 (3.96 in Fig.~\ref{fig:modes}) & $c_2\simeq\sqrt{f_s}\,c_u$ with $c_2\simeq1.4\,c_T$; gives $c_+/c_T=7.1$, $c_-/c_T=1.4$ \\
Atoms $N$ & $(2.3$--$5.8)\times10^6$ & $n_{2D}\,4\pi R^2$, with $n_{2D}$ from Refs.~\cite{Poli2024Excitations2dSS,Ripley2023Planar} \\
Linewidth $\Gamma$ & 1 and 6~s$^{-1}$ & supersolid lifetime~\cite{Sohmen2021Birth}; best measured mode damping~\cite{Tanzi2019SymmetryBreaking} \\
Frequency unit & 7.3~Hz & $c_T/(2\pi R)$ \\
\end{tabular}
\end{ruledtabular}
\end{table*}
 
\begin{table*}[ht!]
\caption{Rydberg-dressed (soft-core) $^{87}$Rb values used in Fig.~\ref{fig:modes}(e)--(h), from the 2D soft-core supersolid at $\Lambda=42.5$ of Ref.~\cite{Poli2024Excitations2dSS}. For now these values are a simulation: no Rydberg-dressed supersolid has been observed yet.}
\label{tab:ryd}
\begin{ruledtabular}
\begin{tabular}{l l l}
Quantity & Value & Source and reasoning \\
\colrule
Platform & soft-core bosons & isotropic cluster crystal over the full shell~\cite{Ciardi2024BubbleSupersolids} \\
$f_s$ & 0.5 & $\simeq0.6$ at the transition by path-integral Monte Carlo~\cite{Macri2013SoftCore} \\
$c_\theta^2/c_T^2$, $2\lambda/n$ & 0.6, 0.14 & density integrated out of the elastic parameters; $g^2/(\chi\mu_e)=0.32$ \\
$B/\mu_e$ & 2.15 (1.44 in Fig.~\ref{fig:modes}) & $(\alpha_{uu}-\alpha_{\rho u}^2/\alpha_{\rho\rho})/\tilde\mu$; gives $c_+/c_T=1.88$, $c_-/c_T=0.73$ \\
Soft-core radius $a_{sc}$ & $0.5~\mu$m & $\simeq0.4~\mu$m in Ref.~\cite{Cinti2010Droplet} \\
Spacing $d$ & $0.73~\mu$m ($R/d=21$) & $d\simeq1.45\,a_{sc}$~\cite{Macri2013SoftCore} \\
$c_T$ & 6.7~mm/s & $c_t\simeq4.6\,\hbar/(M a_{sc})$ \\
Atoms $N$ & $2.8\times10^4$ & areal density from $\Lambda$ with $U_0/k_B=120$~nK~\cite{Cinti2010Droplet} \\
Linewidth $\Gamma$ & 10 and 30~s$^{-1}$ & dressed lifetimes $>100$~ms~\cite{Weckesser2025Dressed}; $\varepsilon_0/(h\gamma)\simeq90$~\cite{Cinti2010Droplet} \\
Frequency unit & 71~Hz & $c_T/(2\pi R)$ \\
\end{tabular}
\end{ruledtabular}
\end{table*}

\end{document}